\documentclass[twocolumn]{aastex631}
\usepackage{graphicx}
\usepackage{natbib}
\usepackage{aas_macros}
\usepackage{wasysym}

\accepted{September 29, 2026}

\shortauthors{Cordiner et al.}

\begin{document}

\title{Temporal Variability of Titan's Middle-Atmospheric Zonal Winds from Southern Fall to Late Winter (2016--2023)}

\correspondingauthor{Martin Cordiner}
\email{martin.cordiner@nasa.gov}


\author[0000-0001-8233-2436]{Martin A. Cordiner}
\affiliation{Astrochemistry Laboratory, NASA Goddard Space Flight Center, 8800 Greenbelt Road, Greenbelt, MD 20771, USA.}
\affiliation{Department of Physics, Catholic University of America, Washington, DC 20064, USA.}

\author[0000-0002-6023-5755]{Yuan Lian}
\affiliation{Aeolis Research, 333 N. Dobson Road, Unit 5, Chandler, AZ 85224, USA.}

\author{Juan M. Lora}
\affiliation{Department of Earth and Planetary Sciences, Yale University, New Haven, CT, USA.}

\author{Nicholas A. Lombardo}
\affiliation{Department of Earth and Planetary Sciences, Yale University, New Haven, CT, USA.}

\author[0000-0002-6023-5755]{Lucy Wright}
\affiliation{School of Earth Sciences, University of Bristol, Wills Memorial Building, Queens Road, Bristol, BS8 1RJ, UK.}

\author[0009-0004-1366-9472]{Jonathon Nosowitz}
\affiliation{Department of Physics, Catholic University of America, Washington, DC 20064, USA.}
\affiliation{Astrochemistry Laboratory, NASA Goddard Space Flight Center, 8800 Greenbelt Road, Greenbelt, MD 20771, USA.}

\author{Richard Cosentino}
\affiliation{Space Telescope Science Institute, 3700 San Martin Drive, Baltimore, MD 21218, USA.}

\author[0000-0003-3108-5775]{Nicholas A. Teanby}
\affiliation{School of Earth Sciences, University of Bristol, Wills Memorial Building, Queens Road, Bristol, BS8 1RJ, UK.}

\author[0000-0001-9540-9121]{Conor A. Nixon}
\affiliation{Astrochemistry Laboratory, NASA Goddard Space Flight Center, 8800 Greenbelt Road, Greenbelt, MD 20771, USA.}

\author{Alexander E. Thelen}
\affiliation{Division of Geological and Planetary Sciences, California Institute of Technology, Pasadena, CA 91125, USA.}

\author{Claire E. Newman}
\affiliation{Aeolis Research, 333 N. Dobson Road, Unit 5, Chandler, AZ 85224, USA.}

\author[0000-0001-6752-5109]{Steven B. Charnley}
\affiliation{Astrochemistry Laboratory, NASA Goddard Space Flight Center, 8800 Greenbelt Road, Greenbelt, MD 20771, USA.}

\author{Eliot F. Young}
\affiliation{Southwest Research Institute, 1301 Walnut Street, Boulder, CO 80302, USA.}

\begin{abstract}
Previous ALMA observations have revealed unexpectedly strong and rapid variations in Titan's high-altitude equatorial zonal winds, which present a challenge to our understanding of this moon's atmospheric dynamics. Here we report further measurements of Titan's zonal wind field at middle atmospheric altitudes $\approx200$--490~km, based on ALMA observations of spatially and spectrally resolved CH$_3$CN emission between 2016--2023 (corresponding to a solar longitude $L_s=81.6^{\circ}$--$156.4^{\circ}$, {spanning Titan's late southern autumn to late winter}). These observations indicate substantial, ongoing, rapid temporal variability of Titan's zonal winds, on timescales as short as $\approx1$ Earth month ($0.9^{\circ}$ in $L_s$). The strongest variability is at (near-)equatorial latitudes where the zonal winds are fastest (up to $240\pm26$~m\,s$^{-1}$), with repeated $\sim40$~m\,s$^{-1}$ fluctuations occurring between $L_s=146.0^{\circ}$--$156.4^{\circ}$.  The observed longer-term trends are qualitatively well-reproduced by two independent, state-of-the-art general circulation models (GCMs), which match the observed decrease in zonal wind speeds over a broad range of latitudes between $L_s=81.6^{\circ}$--$146.0^{\circ}$, followed by a more moderate increase between $L_s=146.0^{\circ}$--$156.4^{\circ}$. The wind speeds at mid-to-high latitudes ($\sim-45^{\circ}$ south) are also typically reasonably well reproduced by the GCMs {(differing by $\lesssim30$\%)}. However, the equatorial wind speeds are up to a factor of $\sim2$ faster than model predictions (with the greatest discrepancy around solstice; $L_s\approx90^{\circ}$). Rapid temporal variability of the retrieved zonal winds indicates the presence of strong atmospheric instabilities that are not well reproduced by models, suggesting a need for future GCM improvements.

\end{abstract}


\section{Introduction}

Titan, the largest moon of Saturn, possesses a thick (1.45 bar) atmosphere composed primarily of N$_2$ ($\approx98$\%) and CH$_4$ ($\approx2$\%), in addition to a suite of trace nitriles, oxygen-bearing molecules and more complex hydrocarbons, discovered by in situ and remote sensing \citep{nix24}. Titan is a slowly rotating body (with a period of 15.9 Earth days), yet it hosts a system of fast zonal winds that encircle the entire moon, the presence of which was initially deduced from cyclostrophic wind balance and stellar occultation observations \citep{fla81,hub93}, and later confirmed (in the altitude range $z=0$--150~km) by in-situ Cassini-Huygens Doppler measurements \citep{bir05}. Stratospheric zonal wind speeds up to $200$~m\,s$^{-1}$ (at $z\sim100$--400~km) were derived indirectly from Cassini Composite Infrared Spectrometer (CIRS) infrared CH$_4$ temperature measurements \citep{fla05,ach08,ach11} using the {gradient} wind equation. {Although the spatial resolution in the wind fields achieved using CIRS temperature data was very good ($\sim$ 10 degrees in latitude and $\sim40$~km in altitude), this technique breaks down at (near-)equatorial latitudes, and was typically only sensitive to Titan's winds below $z\sim500$~km}. The first \emph{direct} measurement of Titan's high-altitude zonal winds (at $z\sim300$--450~km) was made using mm-wave observations with the Plateau de Bure Interferometer \citep{mor05}. \citet{lel19} subsequently used the Atacama Large Millimeter/submillimeter Array (ALMA) to perform more detailed, global zonal wind mapping, extending from the middle-to-upper atmosphere (at altitudes $z\sim300$--1000~km), and discovered an unexpectedly fast, superrotating equatorial jet at the highest, thermospheric altitudes ($z\sim1000$~km). The fast zonal winds traverse the globe on timescales much shorter than a single Titan day, and are thus responsible for the mixing and redistribution of chemical species between day and night hemispheres \citep{cui09,cor19}, as well as helping trap photochemical products at high winter latitudes, within the polar vortex \citep{tea07,tea18}. Persistent zonal winds at lower altitudes are also thought to play a role in shaping Titan's organic-rich `dunes' \citep{lar19}. A more complete, global characterization of the zonal winds is necessary since robust observations do not yet exist across many seasonal transitions.

Recent ALMA observations revealed unexpectedly strong and rapid variations in Titan's high-altitude zonal winds \citep{cor20}. During the relatively short, nine-Earth-month period between August 2016 and May 2017 (between Solar longitude $L_s=82^{\circ}$--$90^{\circ}$, where $L_s=90^{\circ}$ is Titan's northern summer solstice), the thermospheric equatorial wind speed --- probed using Doppler shifted emission lines from the HNC molecule (sounding altitudes $z\sim1000$~km) --- was found to have decreased by a surprising 47\%, from $373\pm6$~m\,s$^{-1}$ to $196\pm2$~m\,s$^{-1}$. At lower, mesospheric altitudes ($z\sim350$~km), a more modest decrease of 27\% was observed, from $254\pm4$~m\,s$^{-1}$ to $185\pm3$~m\,s$^{-1}$.  While Titan general circulation models (GCMs) {have achieved reasonable success in explaining} the basic physical mechanisms leading to atmospheric superrotation \citep{hou95,leb09,new11,lor15,lew23,lia25}, the extreme and rapid high-altitude zonal wind speed variability is not represented by the present generation of models. This may be related to the limitations on the maximum model altitudes ($z\sim500$~km), which inherently exclude mesosphere-thermosphere phenomena. Therefore, although the (intermittent) driving mechanism for the high-velocity thermospheric jet remains unknown \citep{cor20}, it may be related to vertical transport of angular momentum to/from high altitudes (above $\sim500$~km) in the equatorial region. ALMA observations thus highlight new avenues toward improved characterization and understanding of Titan's global circulation system, with implications for theories regarding the climate, zonal winds, and mechanisms for atmospheric momentum transport on Titan that may be applicable to other superrotating atmospheres of rocky planets in our Solar System, and elsewhere in the Galaxy.  

Our present observational knowledge of Titan's equatorial zonal wind field in the stratosphere and above has been limited to a handful of studies \citep{mor05,lel19,cor19,lig24,mar25}. Additional observations are required to more fully characterize the winds as a function of time, to help shed light on the intrinsic cadence (and magnitude) of temporal variability. This will allow us to further investigate the properties of the high-altitude jet(s), and, along with GCM analyses, study the cause of their variability with respect to seasonal (or more rapid) changes in insolation and other atmospheric circulation phenomena. In this article, we present new and archival ALMA observations of CH$_3$CN zonal winds from a total of seven dates between 2017--2023, to characterize the evolution of Titan's wind field in the upper-stratosphere to lower-mesosphere. The resulting latitudinal wind fields are compared against simulations using the Titan Weather Research and Forecasting (TitanWRF) and Titan Atmospheric Model (TAM) GCMs, to provide insights into the physical origin of the winds, and their observed variability.

\section{Observations}
\label{sec:obs}

New observations of Titan were obtained on five epochs between UT 2022-06-10 and 2023-05-09 using the ALMA Band 7 and 9 receivers. The Band 7 observations (four epochs) covered rotational emission from 4 lines of CH$_3$CN, in addition to lines of HNC, HC$_3$N and C$_2$H$_5$CN in the range 349--364 GHz (only the CH$_3$CN data were used in the present study). The Band 9 observations (a single epoch; 2022-09-06) covered 8 lines of CH$_3$CN between 643--662~GHz \citep[see][for further details]{mar25}. For Band 7, the ALMA correlator was set to a spectral resolution of $\Delta{\nu}=61$~kHz (corresponding to a velocity sampling $\Delta{v}=50$--52~m\,s$^{-1}$), while the Band 9 observations were set to $\Delta{\nu}=122$ kHz ($\Delta{v}=55$--57~m\,s$^{-1}$). A summary of the observations is given in Table \ref{tab:obs}.  Also included in the present study are the previously published Titan Band 7 CH$_3$CN data from \citet{lel19} and \citet{cor20}, which were re-reduced and analyzed in the same way as the newly-acquired data, to enable consistent comparison of the results. Those data were obtained at a lower spectral resolution ($\Delta{\nu}=488$ kHz; $\Delta{v}=420$~m\,s$^{-1}$), but covered additional lines of CH$_3$CN (7 in total, which include the 4 lines observed as part of the 2022-2023 data), in the range 349.2--349.5~GHz.  The observed CH$_3$CN spectral line frequencies are given in Table \ref{tab:lines}.

The interferometric data were flagged and calibrated in the Common Astronomy Software Applications (CASA) package \citep{casa22} using standard scripts supplied by the Joint ALMA Observatory. Additional self-calibration of the antenna phases and amplitudes was performed with respect to the Butler-JPL-Horizons 2012 Titan model included with CASA\footnote{https://science.nrao.edu/facilities/alma/aboutALMA/Technology\\/ALMA\_Memo\_Series/alma594/memo594.pdf}, using separate (line-free) continuum regions adjacent to the main spectral windows of interest. 

Prior to imaging, the Titan continuum was subtracted from the interferometric visibility data using a 2nd order polynomial fit to the line-free (pseudo-)continuum regions in each spectral window. The spectral image data were deconvolved (per channel) using the CASA {\tt tclean} H{\"o}gbom algorithm with Briggs weighting and stopping at a threshold of twice the RMS noise.  A Briggs \texttt{robust} parameter of 0.5 was used for all epochs apart from 2023-05-09, for which $robust = 1.0$ was used, to produce an improved RMS noise, at the expense of spatial resolution. Even using $robust = 1.0$, the 2023-05-09 data have a significantly smaller beam size than the other epochs, as a result of the more extended ALMA array configuration used for these observations. 

\begin{figure}[h!]
\centering
\includegraphics[width=7cm]{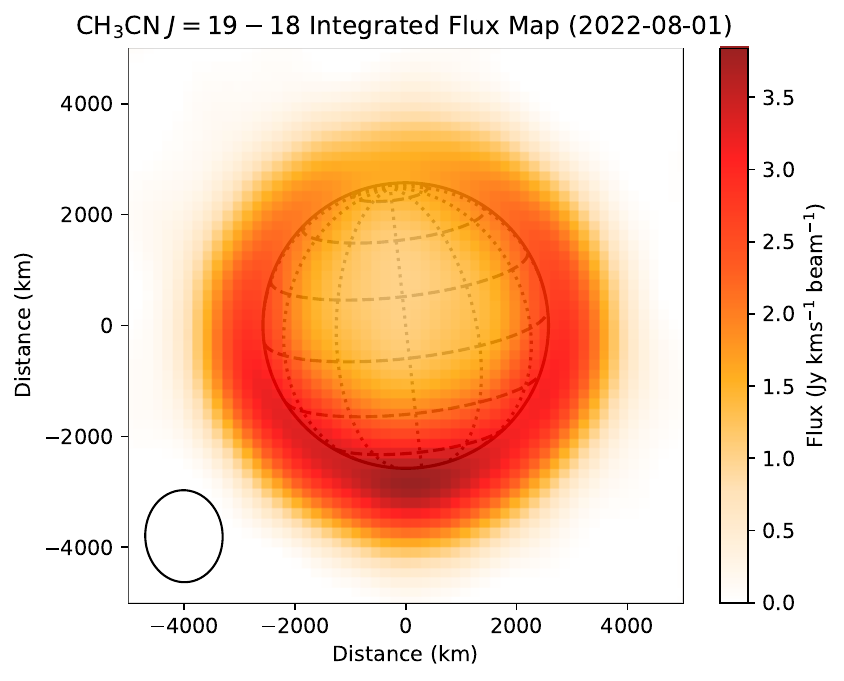}
\includegraphics[width=\columnwidth]{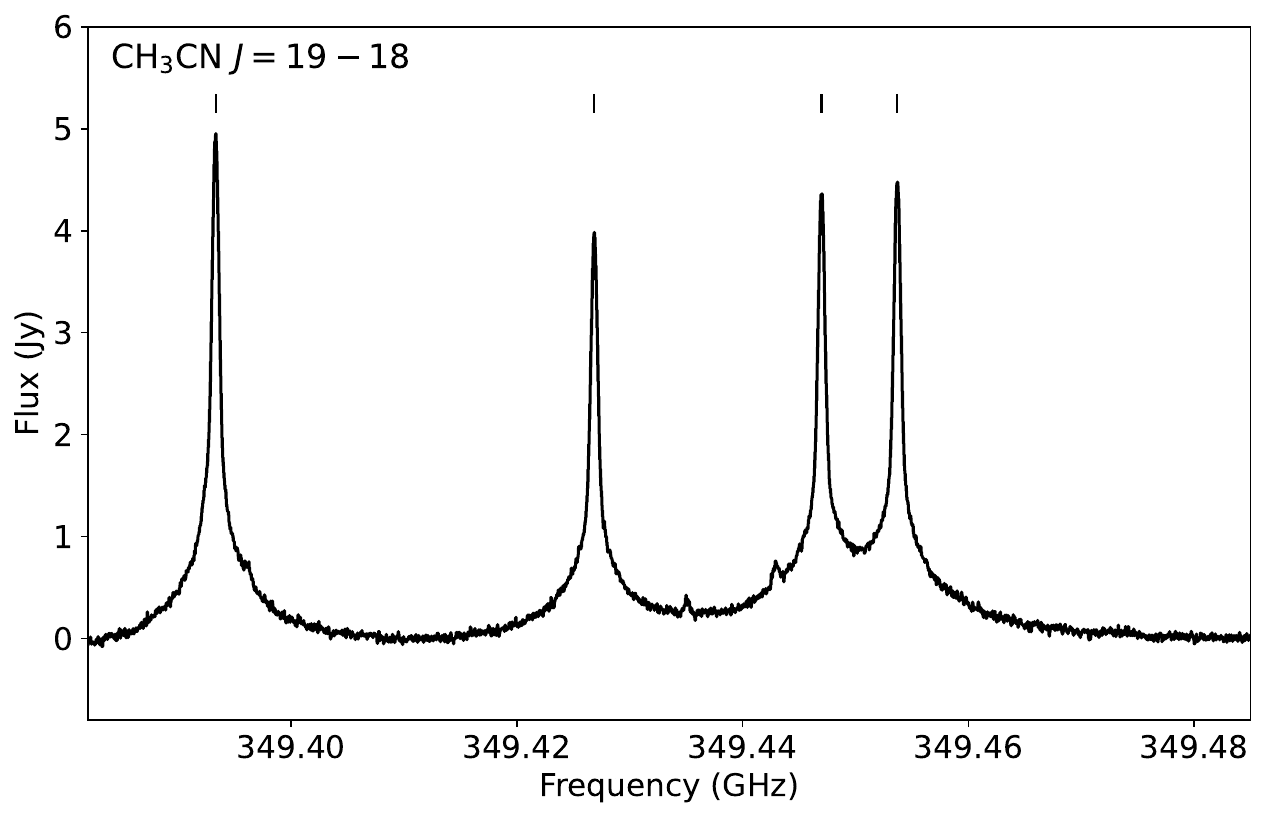}
\caption{Continuum-subtracted ALMA observations of Titan on 2022-08-01, showing the spectrally integrated line flux map for the CH$_3$CN $J=19-18$ band (left panel) and corresponding, disk-integrated spectrum (right panel), summed over a $1.5''$-diameter (9,700 km) circular aperture centered on Titan. The FWHM of the spatial resolution element (ellipse) is shown in the lower left of the map. The four strongest lines in the spectrum (marked with ticks) correspond to the ($J_K=19_{K}$--$18_{K}$; $|K|=0$--3) transitions of CH$_3$CN, while the weaker lines (at 349.396~GHz, 349.435~GHz and 349.442~GHz) originate from C$_2$H$_5$CN emission. \label{fig:almadata}}
\end{figure}

Image restoration was performed using the smallest common elliptical Gaussian beam for each measurement set ($\theta_B$; see Table \ref{tab:obs}). For plotting purposes, the data were projected onto a spatial grid in the plane of the sky, according to Titan's geocentric distance ($\Delta$). As a demonstration of the typical data quality, the spectrally integrated (continuum subtracted) line flux map and disk-integrated spectrum for the 2022-08-01 epoch are shown in Figure \ref{fig:almadata}.

\begin{table*}
\caption{Summary of ALMA Observations and Target Geometry \label{tab:obs}}
\hspace{0cm}\begin{tabular}{cccccccccc}
\hline
UT Date & $L_s$ & Freq.  & $\Delta{\nu}$ & $\theta_B$ & $\Delta$ & NP Ang. & Sub-lat. & $\theta_{atm}$ & $\phi_{atm}$  \\
      &  ($^{\circ}$) &  (GHz) & (kHz)   & ($''$) &   (au) & ($^{\circ}$) & ($^{\circ}$)& ($^{\circ}$) & ($^{\circ}$)\\
\hline
2016-08-19 & 81.6  & 349       & 488  & $0.20 \times 0.15$&   9.8 & $-3.4$& 26.1  & 1.1&3.9\\
2017-05-08 & 89.5  & 349       & 488  & $0.23 \times 0.15$&   9.2 & $-5.3$& 26.4  & 2.5&3.3\\
2022-06-10 & 146.0 & 349       & 61   & $0.30 \times 0.21$&   9.4 & $-6.3$& 12.2  & 4.1&$-0.3$\\
2022-08-01 & 147.6 & 349       & 61   & $0.23 \times 0.20$&   8.9 & $-6.4$& 13.3  & 4.1&$-0.1$\\
2022-09-06 & 148.7 & 643,662   & 122  & $0.22 \times 0.17$&   8.9 & $-6.5$& 14.5  & 4.1&$0.0$\\
2023-04-12 & 155.6 & 349       & 61   & $0.23 \times 0.18$&  10.4 & $-5.9$& 8.5   & 3.9&$-0.8$\\
2023-05-09 & 156.4 & 349       & 61   & $0.14 \times 0.13$&  10.1 & $-5.8$& 7.6   & 3.9&$-1.0$\\\hline
\end{tabular}

\hspace{2.7cm}\parbox{0.74\textwidth}{\footnotesize 
Notes --- $L_s$ is the Saturnian solar longitude (which corresponds to the Titan season); $\Delta{\nu}$ is the frequency resolution; $\theta_B$ is the elliptical ALMA beam FWHM dimensions; $\Delta$ is the Earth-Titan distance; NP Ang. is the angle of Titan's north pole (projected in the plane of the sky), measured clockwise from celestial north; Sub-lat. is the sub-observer latitude; $\theta_{atm}$ and $\phi_{atm}$ are the atmospheric polar tilt components relative to Titan's solid-body pole (in the plane of the sky, and out of the plane of the sky, respectively).\\
}
\end{table*}

\begin{table}
\caption{Molecular Line Data \label{tab:lines}}
\hspace{-0.7cm}\begin{tabular}{lcccc}
\hline\hline
Molecule & Transition & Frequency & $E_u$ \\
         & ($J'_{K'}$--$J''_{K''}$)  & (MHz)  & (K)\\
\hline  
   CH$_3$CN & $19_6$--$18_{-6}$ & 349212.3109(2) & 425\\
   CH$_3$CN & $19_{-6}$--$18_6$ & 349212.3109(2) & 425\\
   CH$_3$CN & $19_5$--$18_5$ & 349286.0060(2) & 346\\
   CH$_3$CN & $19_4$--$18_4$ & 349346.3430(2) & 282\\
   CH$_3$CN & $19_3$--$18_{-3}$ & 349393.2973(2) & 232\\
   CH$_3$CN & $19_{-3}$--$18_3$ & 349393.2973(2) & 232\\
   CH$_3$CN & $19_2$--$18_2$ & 349426.8499(2) & 196\\
   CH$_3$CN & $19_1$--$18_1$ & 349446.9869(2) & 175\\
   CH$_3$CN & $19_0$--$18_0$ & 349453.7001(2) & 168\\[2mm]
   CH$_3$CN & $35_{-3}$--$34_3$ & 643159.6935(2) & 620\\
   CH$_3$CN & $35_3$--$34_{-3}$ & 643159.6935(2) & 620\\
   CH$_3$CN & $35_2$--$34_2$ & 643220.8920(3) & 585\\
   CH$_3$CN & $35_1$--$34_1$ & 643257.6211(3) & 563\\
   CH$_3$CN & $35_0$--$34_0$ & 643269.8658(3) & 556\\
   CH$_3$CN & $36_{-3}$--$35_3$ & 661496.8384(10) & 652\\
   CH$_3$CN & $36_3$--$35_{-3}$ & 661496.8384(10) & 652\\
   CH$_3$CN & $36_2$--$35_2$ & 661559.7344(3) & 616\\
   CH$_3$CN & $36_1$--$35_1$ & 661597.4823(3) & 595\\
   CH$_3$CN & $36_0$--$35_0$ & 661610.0668(10) & 588\\
\hline
\end{tabular}\\
\vspace{-3mm}
\begin{center}\parbox{0.39\textwidth}{Notes --- Transition frequencies and upper-state energy levels ($E_u$) are from the CDMS catalogue \citep{mul01}. Uncertainties on the trailing frequency digits are given in parentheses.}\end{center}

\end{table}

\section{Zonal Wind Retrievals}
\label{sec:fits}

We followed a similar strategy to \cite{cor20} to retrieve the latitudinal zonal wind velocity profile on each date, based on Doppler shifts extracted from analytic functional fits to the observed spectral line profiles. The line-fitting method employed by \cite{lel19} and \cite{cor20} was updated to use the Moffat function \citep{mof69,lig24} instead of a Gaussian, which was found by \cite{lig24} to provide a significantly better fit to the CH$_3$CN lineshapes than Gaussian or Voigt functions, especially in the presence of pressure-broadened Lorentzian line wings:

\begin{equation}
    I(\nu) = I_0 \left[ 1 + \left( \frac{\nu-\nu_0}{\alpha} \right)^2 \right]^{-\beta}
\end{equation}

where $I_0$ is the peak line strength, $\nu_0$ is the peak offset, and $\alpha$ and $\beta$ parameterize the widths of the line core and wings, respectively. We found that the Moffat function was able to reproduce our observed line profiles to surprisingly high accuracy (within the noise), resulting in a more accurate determination of the line Doppler shift than the previously applied Gaussian fitting methods.

\begin{figure}[h!]
\centering
\includegraphics[width=8cm]{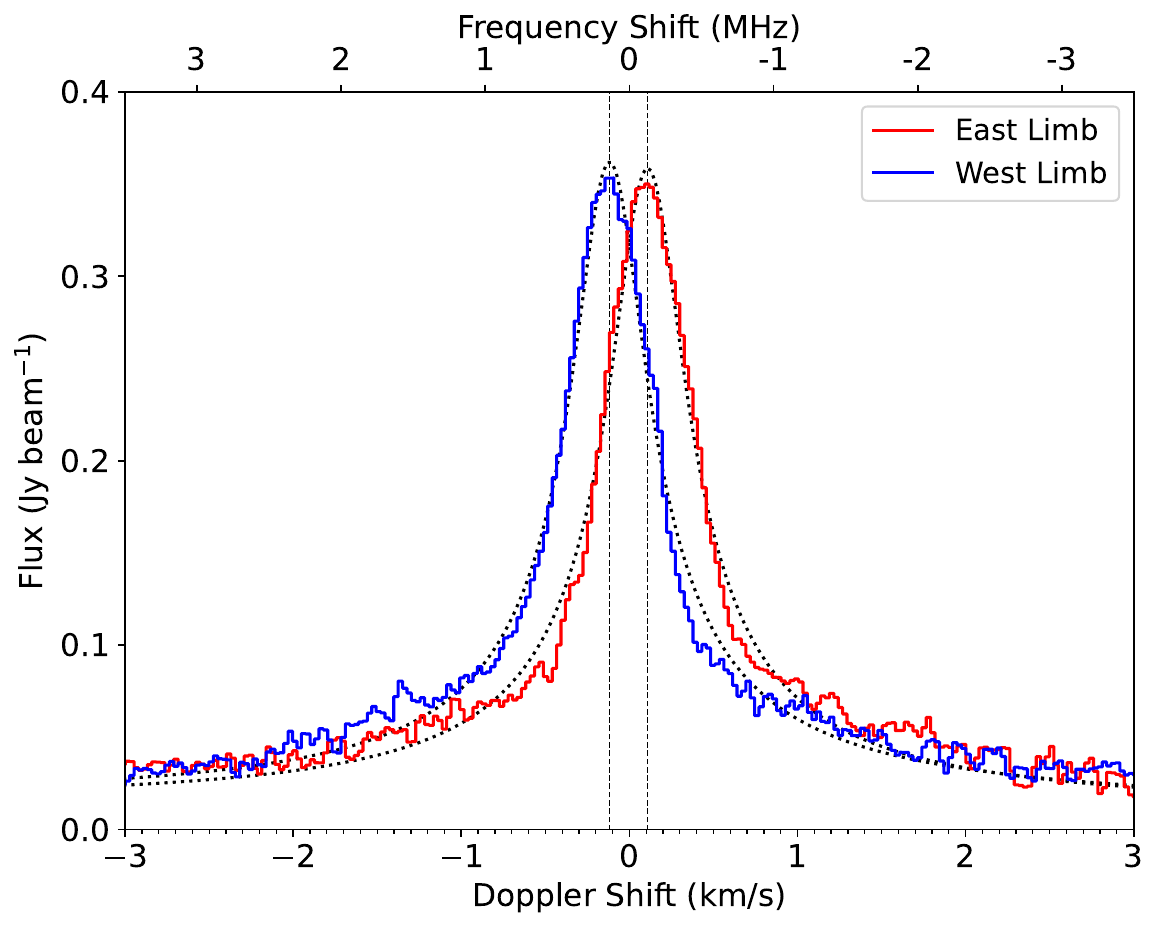}
\caption{Demonstration of Moffat function fits to the CH$_3$CN ($J_K=19_2$--$18_2$) spectra obtained on 2022-06-10. The continuum-subtracted ALMA observations were extracted at the east and west equatorial limbs (red and blue curves, respectively). Moffat fits used to derive the Doppler velocities are overlaid as dotted black curves; vertical dashed lines indicate the best-fitting Doppler shifts associated with each fit. \label{fig:fits}}
\end{figure}

For each date, and for each image pixel, all CH$_3$CN spectral lines within the observed frequency range were fitted simultaneously, using the LMFIT (Levenberg--Marquardt) least-squares minimization routine \citep{new16a}. All four parameters of the Moffat function were allowed to vary independently for each line, apart from the velocity offset parameter, for which a common (variable) value was used for all CH$_3$CN lines. {A demonstration of the typical quality of the spectral line fits is shown for the east and west equatorial limb positions in Figure \ref{fig:fits}.} The uncertainty on the Doppler shift for each pixel was derived from the diagonal elements of the covariance matrix. The resulting line-of-sight Doppler velocity maps (relative to Titan's inertial rest frame) are shown in Figures \ref{fig:maps1} and \ref{fig:maps2}; their associated uncertainty maps are shown in Appendix A, Figures \ref{fig:errmaps1} and \ref{fig:errmaps2}.

For the 2022-09-06 epoch, a significant ($\approx+30$ m\,s$^{-1}$) offset in the spectral line Doppler shifts was identified at the center of Titan's disk (and consistently, along the length of the sky-projected polar axis).  This is attributed to an error in the cable routing of the ALMA rubidium maser at the time of these observations (which affected only the 2022-09-06 data). This error is known to have introduced erroneous (and unknown) frequency shifts into similarly-affected ALMA data, accompanied by a slight broadening of the spectral lines by a few tens of meters per second (R. Loomis, private communication).  The magnitude of the shift was calculated by taking the average of the fitted line-of-sight velocities across all pixels lying inside Titan's limb. This ``zero-point shift'' was then subtracted from the individual pixel Doppler shifts for the subsequent data analysis steps. No significant evidence for {longitudinal} asymmetry in Titan's zonal winds was reported by the prior ALMA studies of \cite{lel19}  and \cite{cor20}. Therefore, our ALMA data analysis inherently assumes that the zonal winds are symmetric with respect to Titan's sky-projected polar axis. 

To determine the altitudinal sensitivity of our ALMA zonal wind measurements, we used the Non-linear optimal Estimator for MultivariatE spectral analySIS (NEMESIS; \citealt{irw08}) to generate detailed radiative transfer models for each set of CH$_3$CN observations. First, the calibrated ALMA data cubes were re-imaged in the same way as described in Section \ref{sec:obs}, but without subtracting the Titan continuum emission. A similar modeling strategy to \cite{lel19} was then implemented, extracting and modeling the CH$_3$CN spectra observed at the intersection of Titan's eastern limb with the equator, which lies close to the location of the limb emission peak.  These limb spectra were fitted by optimizing an initial \textit{a priori} CH$_3$CN vertical abundance profile obtained from the best-fitting retrieval results of \cite{lel19}.

Titan's disk (plus atmosphere) was divided into 35 annuli corresponding to emission angles between 0--75 degrees, measured from the top of our model atmosphere ($z=1200$~km), and therefore covering emission from the center of Titan up to a tangential altitude of 1050 km (emission was found to be negligible above 1000 km). {Since contemporaneous temperature data were not available for most of our observational epochs, the 2019 atmospheric temperature profile was adopted from \cite{the20}, as a reasonable approximation}. The ALMA fluxes ($S_{\nu}$) for each epoch were scaled by a constant factor in the range 0.93--1.02, to produce a match between the modeled and observed continuum levels. This scaling mitigates errors on the ALMA flux scale (which are typically $\lesssim10$\%, due to uncertainty in the flux of the amplitude calibration quasar), as well as accounting for small errors in the assumed (\textit{a priori}) temperature around the tropopause, which determines the model continuum level. The CH$_3$CN abundance profile was continuously retrieved at each altitude (with a vertical smoothing length of one atmospheric scale height) to obtain a fit to the observed spectral line profiles. The resulting vertical abundance profiles are shown in Appendix B, Figure \ref{fig:vmr}.  Temporal variability is evident in the stratospheric and mesospheric CH$_3$CN abundances, but only at the 2--3 sigma level, so may not be significant. 

The retrieved CH$_3$CN abundance profiles formed the basis of our ``wind contribution function'' calculations, which were obtained from the corresponding NEMESIS functional derivatives ($dQ/d\log(x)$)$_{\nu}$ of flux ($Q$) with respect to the logarithmic abundance ($x$), as a function of frequency ($\nu$). The $dQ/d\log(x)$ values for each epoch were convolved with the gradient of the (continuum-subtracted) model spectral line profile, to provide weighting according to the contribution of each frequency to the retrieved line Doppler shift (see \emph{e.g.}, \cite{lel19}). The weighted wind-contribution functions were then peak-normalized, and convolved with a Gaussian kernel of standard deviation $\sigma=30$~km, to smooth out minor artifacts related to the discrete emission-angle sampling. The resulting contribution functions are plotted in Appendix B, Figure \ref{fig:contrib}.  Apart from the Band 9 2022-09-06 data, the corresponding peak emission altitudes are in the range 247--261~km, with an average of 251 km. The contribution function for 2022-09-06 peaks at 271 km, but is excluded from this range due to additional, unquantified uncertainty on its vertical abundance profile that may have resulted from the rubidium maser cabling problem identified earlier in this section.  The $\pm68.3$\% percentile range about the peak of the average contribution function curve lies between 200--490~km, which corresponds approximately to the $1\sigma$ range of the vertical emission distribution. {We investigated the impact of uncertainties in the atmospheric temperature profile on the contribution functions by re-running the CH$_3$CN abundance retrievals using disk-averaged CO temperature data from 2017 and 2022 \citep{the22,the24}. Although significant differences in the retrieved CH$_3$CN abundances were identified in the lower stratosphere ($z=80$--150~km), line pressure broadening renders our results largely insensitive to winds in that region. Small differences in the retrieved CH$_3$CN abundances above $z\sim150$~km were found to lie within the $1\sigma$ retrieval uncertainties, and therefore may not be significant. Nevertheless, these tests show that the uncertainties on the wind contribution functions in Figure \ref{fig:contrib} as a result of unaccounted-for inter-annual temperature variability may be up to $\pm\sim$ a few percent.}

Due to the effects of beam-smearing, combined with Titan's axial tilt, extraction of the intrinsic zonal wind field requires deconvolution of the Doppler wind map data. We therefore followed a similar method to \cite{cor20} to model the Doppler velocity maps (Figures \ref{fig:maps1} and \ref{fig:maps2}), and extract latitudinal wind speed profiles for each date using a simplified, 3D radiative transfer model. The {\tt atmWindMapGaussFit} code of \cite{cor20} was adapted to allow a continuously variable wind speed ($v$) as a function of latitude ($l$), instead of the Gaussian profile previously assumed by those authors. The $v(l)$ profile was generated by joining a set of $i$ points ($v_i(l_i)$, equally spaced in $l$) by a segmented linear function. This function was then smoothed by a Gaussian function of FWHM equal to the spacing between the points. A sigmoid-type exponential taper was applied at the ends of the $v(l)$ profile (with an exponential smoothing scale length of 1$^{\circ}$), to ensure a velocity of zero at the poles. Each velocity point $v_i$ was then optimized using LMFIT, to obtain the least-squares solution and associated error estimates. The error estimates take into account the covariance of the pixels within each ALMA beam, by dividing each of the (model minus data) LMFIT residuals by the number of unmasked pixels per beam. A circular mask of radius 3500~km was used for all datasets, to exclude the noisiest pixels toward the image edges. Further improvements to the \cite{cor20} methodology included adopting an atmospheric emission model based on the full, retrieved CH$_3$CN VMR profiles (Figure \ref{fig:vmr}) on each date (instead of the coarsely parameterized versions used by \citealt{cor20}). In addition, our updated modeling strategy now also takes into account pressure broadening of the CH$_3$CN spectral lines as a function of altitude and temperature, using the Lorentzian half-width line-broadening coefficients from \citet{dud15}. The resulting zonal wind velocity profiles ($v(l)$) are relative to Titan's surface (\emph{i.e.} with the line-of-sight component of Titan's solid-body rotation subtracted, which amounts to 12~m\,s$^{-1}$ at the equator).

The zonal wind modeling procedure explicitly accounts for the effect on the line-of-sight wind speeds of Titan's polar axis tilt with respect to the plane of the sky (the sub-observer latitude and north pole rotation angle, clockwise in the plane of the sky, are given in Table \ref{tab:obs}). However,  based on 13 years of Cassini CIRS temperature data, \citet{wri25} deduced that the symmetry axis (pole) of Titan's zonal winds ($\tilde{w}$) is offset from that of the solid body, and the degree of offset varies as a function of time. We tested our models for inclusion of this effect and found an improvement in the quality of the fits to the observed Doppler maps by assuming a north polar wind-axis tilt of $4^{\circ}$, fixed in an inertial frame directed $120^{\circ}$ west of the Titan–Sun vector at the northern spring equinox. This is the mean tilt determined by \citet{wri25} over the duration of the Cassini mission. The associated angular offsets of $\tilde{w}$ from the solid-body pole are $\theta_{atm}$ (in the plane of the sky) and $\phi_{atm}$ (out of the sky-plane), which are given for each epoch in Table \ref{tab:obs}. Changes in the retrieved $v(l)$ profiles following the inclusion of these wind-axis corrections to the model are all smaller than the $1\sigma$ uncertainties on $v(l)$. Nevertheless, the inclusion of the atmospheric tilt led to a significant improvement in the (model minus observation) fit residuals, particularly in the vicinity of the poles.

\begin{figure}[h!]
\centering
\includegraphics[width=\columnwidth]{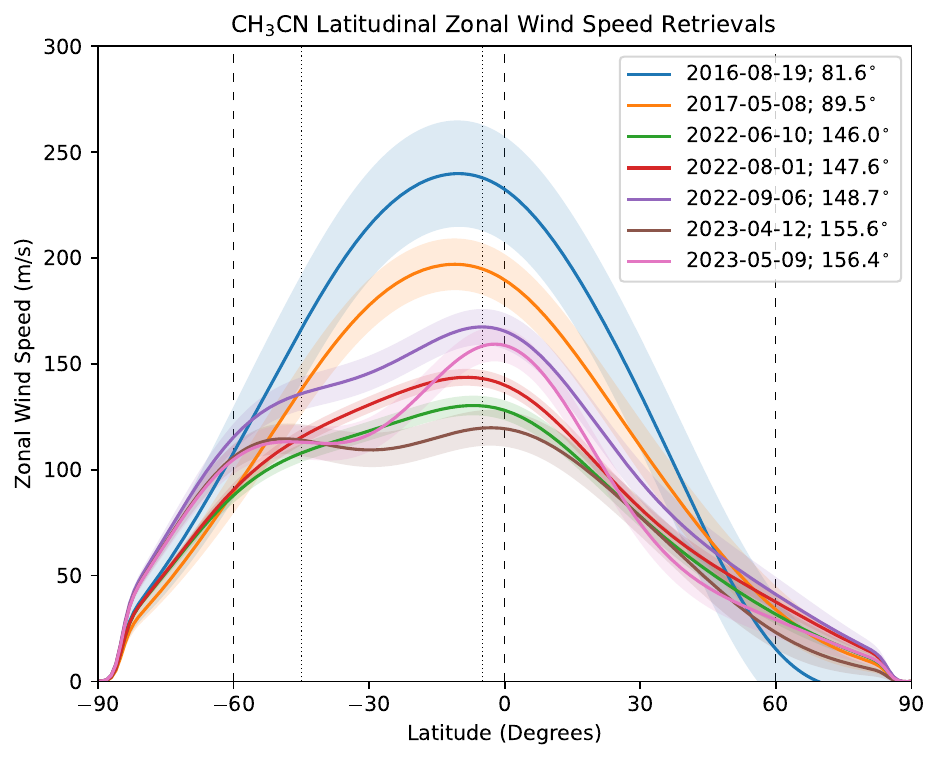}
\caption{Observed zonal wind velocity profiles ($v(l)$) and $1\sigma$ error envelopes retrieved from our ALMA CH$_3$CN observations on seven epochs, using a continuously variable parameterization with latitudinal sampling of 30$^{\circ}$. These wind profiles are primarily sensitive to zonal winds in the altitude range 200--490~km. $L_s$ values for each observation date are also shown in the legend. Vertical dashed lines indicate latitudes $l=-60^{\circ}$, $0^{\circ}$, and $60^{\circ}$; dotted lines indicate $l=-45^{\circ}$ and $-5^{\circ}$, to indicate the latitudes plotted in Figure \ref{fig:windtime}. \label{fig:windlat}}
\end{figure}

To maximize the information content in the $v(l)$ retrievals, we experimented in fitting the observations with latitudinal profiles incorporating different numbers of variable latitudinal points $i$. Seven latitudinal points (with $\Delta_l=30^{\circ}$ separation) corresponds to a distance of $0.2''$ on the sky in the equatorial region, which is comparable to the spatial resolution of the majority of the ALMA data (Table \ref{tab:obs}). This latitudinal sampling was found to produce a visually good fit to the observed Doppler maps, while still allowing the retrieved profile to be well constrained by the data (\emph{i.e.} keeping the resulting $v(l)$ errors reasonably small with respect to variations in the retrieved profile). The resulting $v(l)$ profiles (and associated error envelopes) are shown in Figure \ref{fig:windlat}.

\begin{figure}
\begin{center}
 \includegraphics[width=0.8\columnwidth]{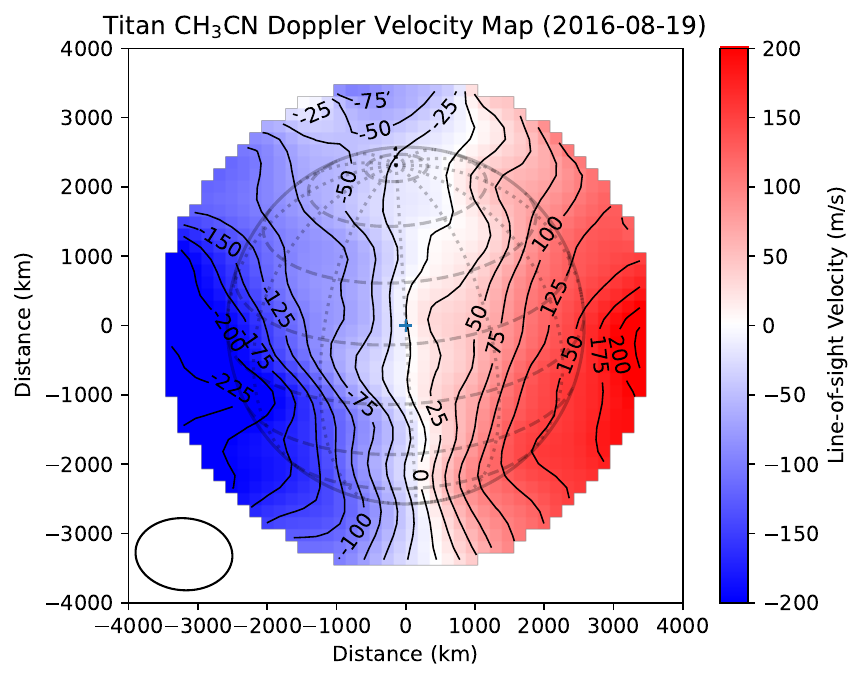}\\
 \includegraphics[width=0.8\columnwidth]{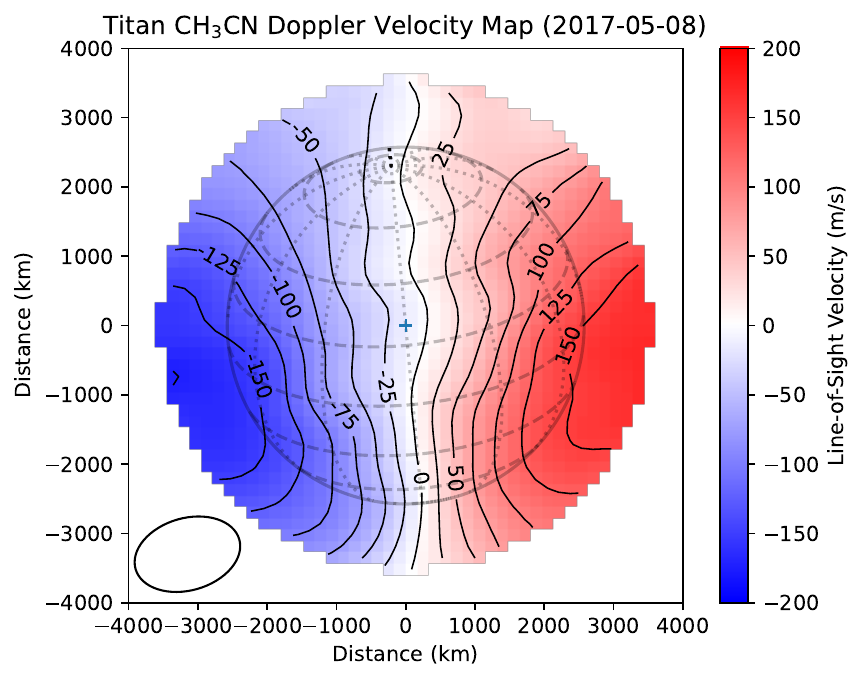}\\
 \includegraphics[width=0.8\columnwidth]{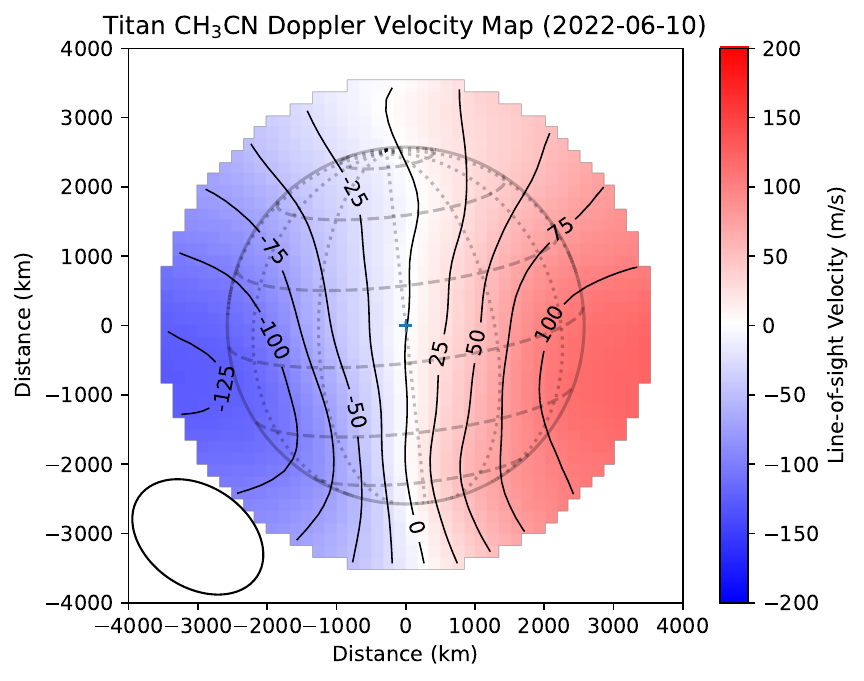}\\
 \includegraphics[width=0.8\columnwidth]{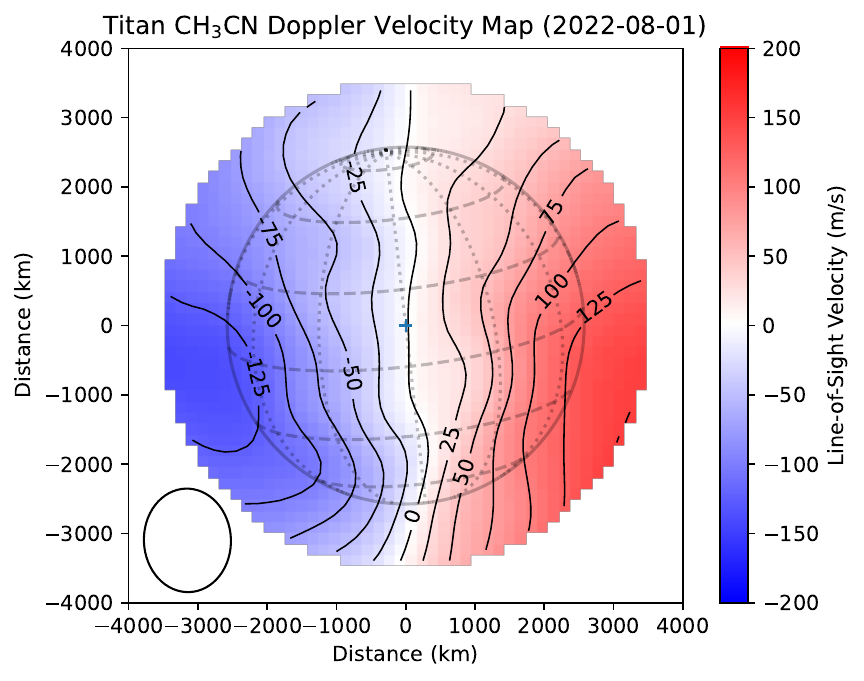}\\
 \caption{Color maps showing line-of-sight CH$_3$CN Doppler shifts for four epochs (2016-08-19 to 2022-08-01). Titan's wireframe grid is shown for each epoch, with the spatial resolution (elliptical beam FWHM) indicated lower left. Velocity contours (labeled in units of m\,s$^{-1}$) are also overlaid.}
   \label{fig:maps1}
\end{center}
\end{figure}

\begin{figure}
\begin{center}
 \includegraphics[width=0.8\columnwidth]{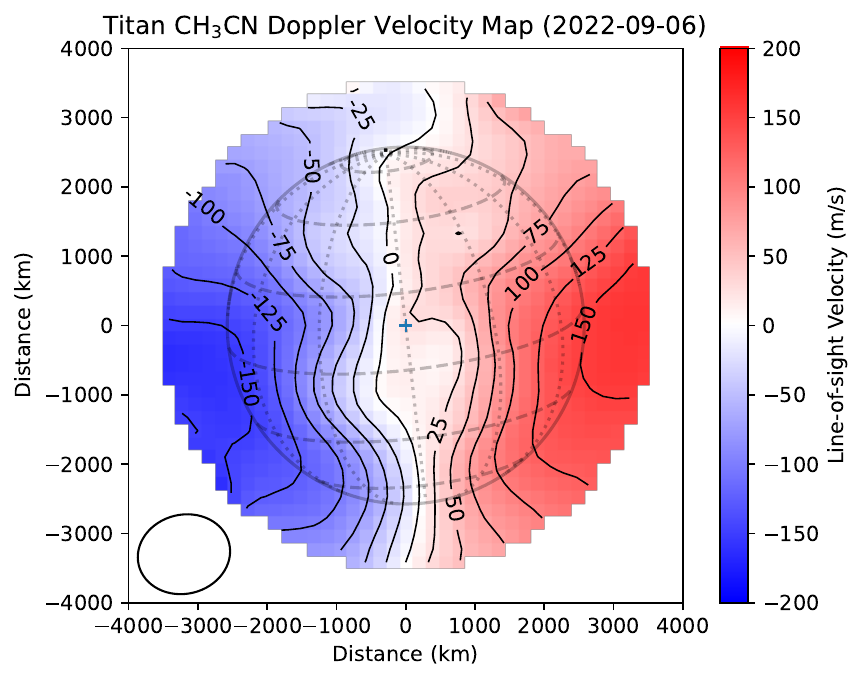}\\
 \includegraphics[width=0.8\columnwidth]{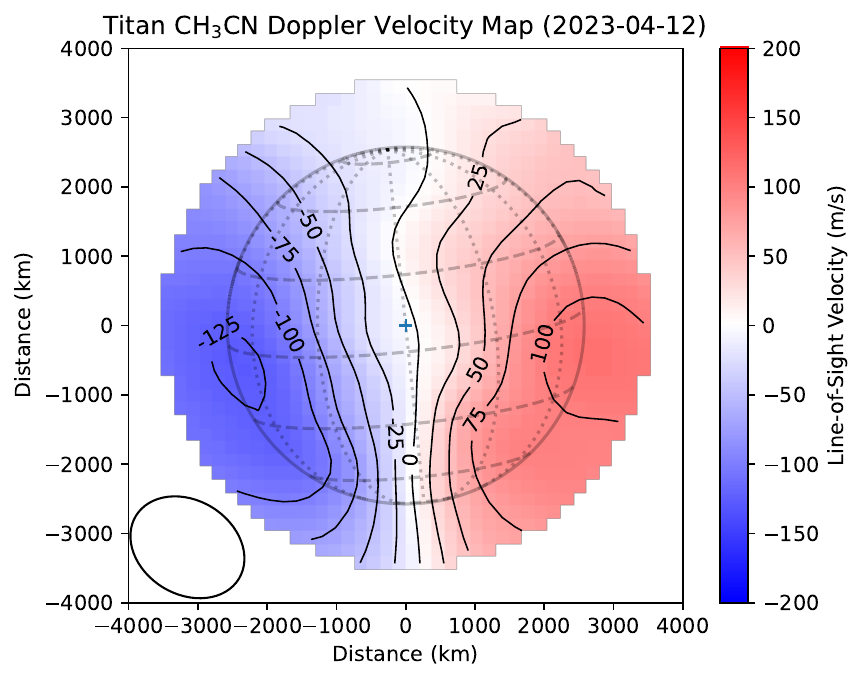}\\
 \includegraphics[width=0.8\columnwidth]{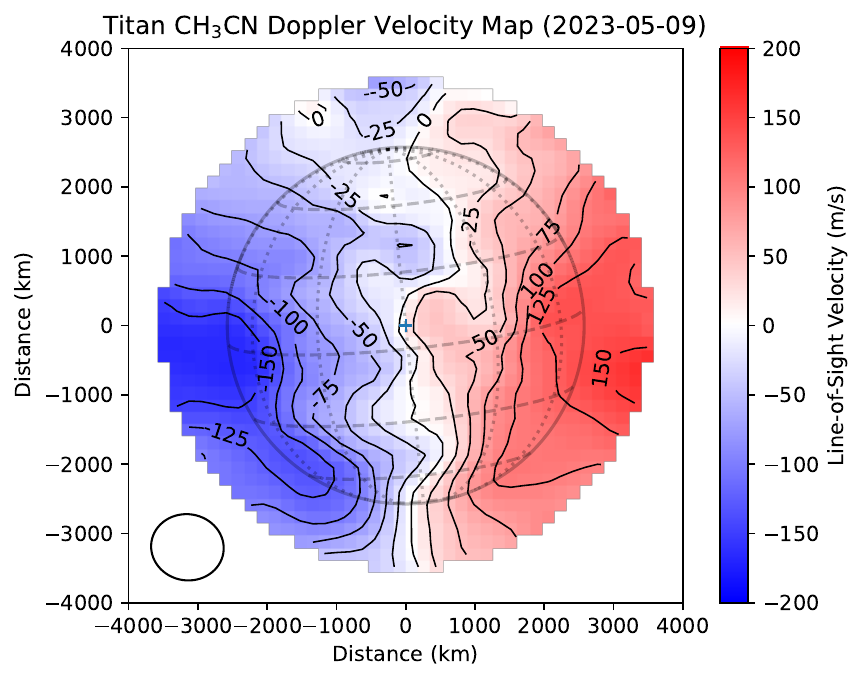}\\
 \caption{Same as Figure \ref{fig:maps1}, for epochs 2022-09-06 to 2023-05-09. Note --- the 2023-05-09 epoch shows increased noise due to a smaller beam size.}
   \label{fig:maps2}
\end{center}
\end{figure}

\section{Properties of the Observed Zonal Wind Patterns}

The retrieved zonal wind speeds show significant variations across all observed epochs. From 2016-08-19 to 2017-05-08 ($L_s=82^{\circ}$--$90^{\circ}$), the change in peak velocity (from $240\pm26$~m\,s$^{-1}$ to $197\pm12$~m\,s$^{-1}$) is comparable to, albeit somewhat less than, that determined by \citet{cor20}, who derived a change from $254\pm4$~m\,s$^{-1}$ to $185\pm3$~m\,s$^{-1}$, for the same period.  The \citet{cor20} study used a Gaussian approximation to the spectral line and $v(l)$ profiles, combined with a simplified radiative transfer model (not taking into account pressure broadening or the detailed CH$_3$CN vertical profile). The earlier study also neglected Titan's solid body rotation (focusing instead on line-of-sight winds relative to Titan's mean inertial frame), so our present wind retrievals should be considered an improvement over the previous results.  In the present study, although our latitudinal resolution ($30^{\circ}$) is admittedly still rather coarse, the resulting $v(l)$ profiles are better able to account for intrinsic latitudinal variations in the wind speeds than the Gaussian $v(l)$ fitting method of \citet{cor20}, as it removes the assumption of latitudinal symmetry.  For the 2016-08-19  and 2017-05-08 epochs, this improvement in the retrieval method manifests as a skewing of the $v(l)$ profiles toward southern (negative) latitudes, which could not be reproduced using a Gaussian $v(l)$ profile. The skewing is even more apparent for the 2022-06-10 and 2022-08-01 epochs, where a statistically significant shoulder appears around $l\approx-50^{\circ}$ with respect to the main peak (at $l\approx-10^{\circ}$). This shoulder is consistent with strong zonal winds in the stratosphere occurring at mid-to-high latitudes in the winter hemisphere, as previously predicted by GCMs \citep{new11,leb12,lor15,lom23b}, and detected in thermally-derived wind fields using Cassini CIRS CH$_4$ data \citep{ach08,ach11,vin20,sha20,wri25}. 

The detailed behavior of Titan's stratospheric to lower-mesospheric high-latitude jet as a function of time has already been revealed by the aforementioned thermal-wind studies. However, {the solution of the gradient wind equation becomes numerically unstable near the equator, exacerbated by the thickness of Titan's atmosphere \citep{fla05}, leading to an incomplete picture of the wind field between latitudes $l\approx\pm10^{\circ}-30^{\circ}$ (depending on altitude)}. General circulation models show consistently fast ($\gtrsim100$~m\,s$^{-1}$), moderately temporally-variable wind speeds at equatorial latitudes, accompanied by stronger, more variable winds at higher (winter) latitudes, which are due to vertical and hemispheric shifts of the jet core as a consequence of seasonal forcing \citep{lom23b}. The ALMA data presented here span $L_s=82^{\circ}$--$156^{\circ}$ --- almost a complete season, from the period just before southern winter solstice, to $24^{\circ}$ before the spring equinox. In contrast to the GCMs, our observations show consistently higher wind speeds around the equator than at mid-to-high latitudes. Our ALMA data also reveal for the first time, strong/ongoing temporal variability at equatorial latitudes over timescales as short as $0.9^{\circ}$ in $L_s$ (27 Earth days), in addition to the smaller changes at mid-to-high latitudes. 

The level of detail in our retrieved zonal wind profiles is inherently limited by the resolution of the ALMA observations, so independent contributions to the $v(l)$ profiles from specific latitudes and altitudes are difficult to pinpoint. Nevertheless, our data are consistent with the presence of fast zonal winds spanning a broad range of latitudes at all epochs, and the relative speeds of the low and high-latitude winds are seen to vary as a function of time. A bilobate $v(l)$ profile is particularly apparent for the 2022-09-06, 2023-04-12 and 2023-05-09 epochs, from which we deduce the presence of distinct contributions from a band of near-equatorial zonal winds (peaking between $l=0^{\circ}$ to $-5^{\circ}$), and a higher latitude band (between $-40^{\circ}$ to $-60^{\circ}$). From the heights of the two lobes in Figure \ref{fig:windlat}, the equatorial and high-latitude southern-hemisphere winds both show significant temporal variability in their observed absolute speeds, in addition to some significant variability between the relative speeds of the two lobes (see also Figure \ref{fig:windlat}). 

\begin{figure}[h!]
\centering
\includegraphics[width=0.8\columnwidth]{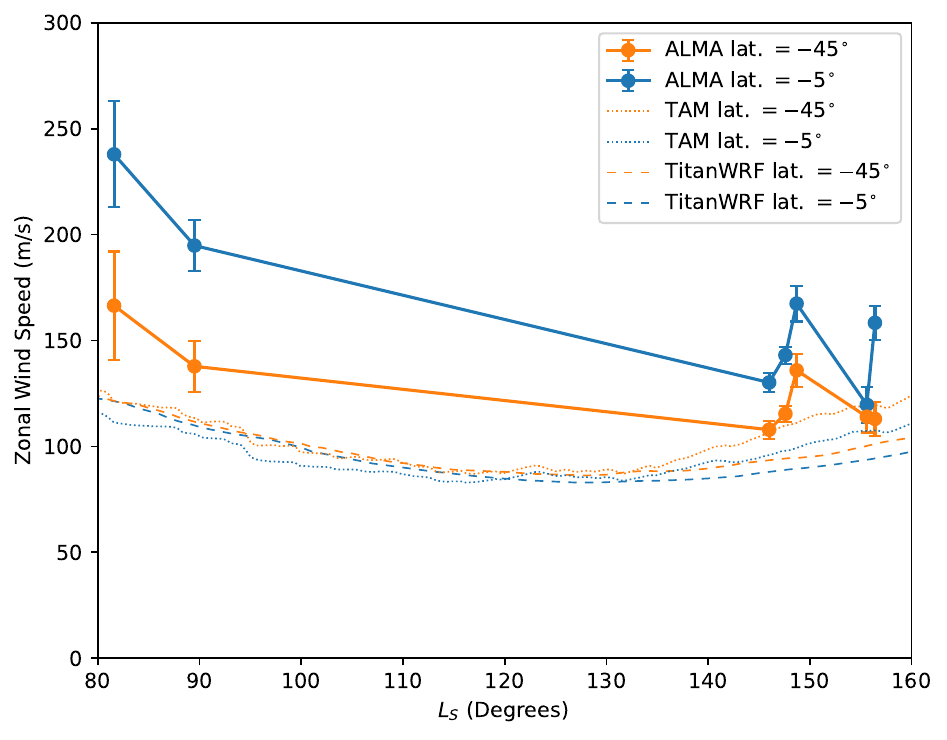}\\
\includegraphics[width=0.8\columnwidth]{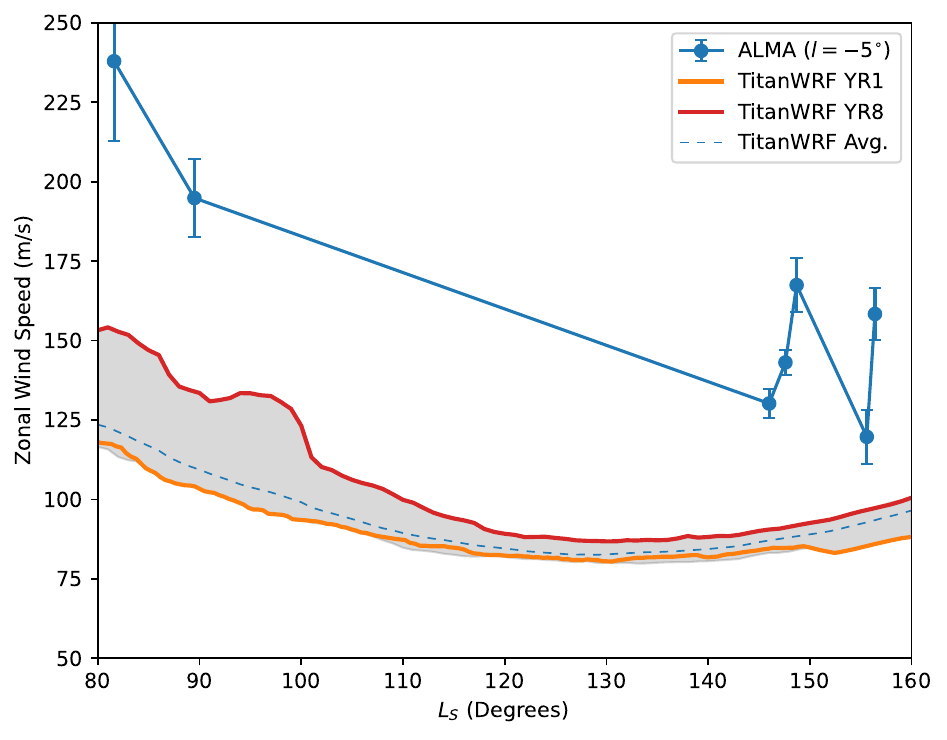}\\
\includegraphics[width=0.8\columnwidth]{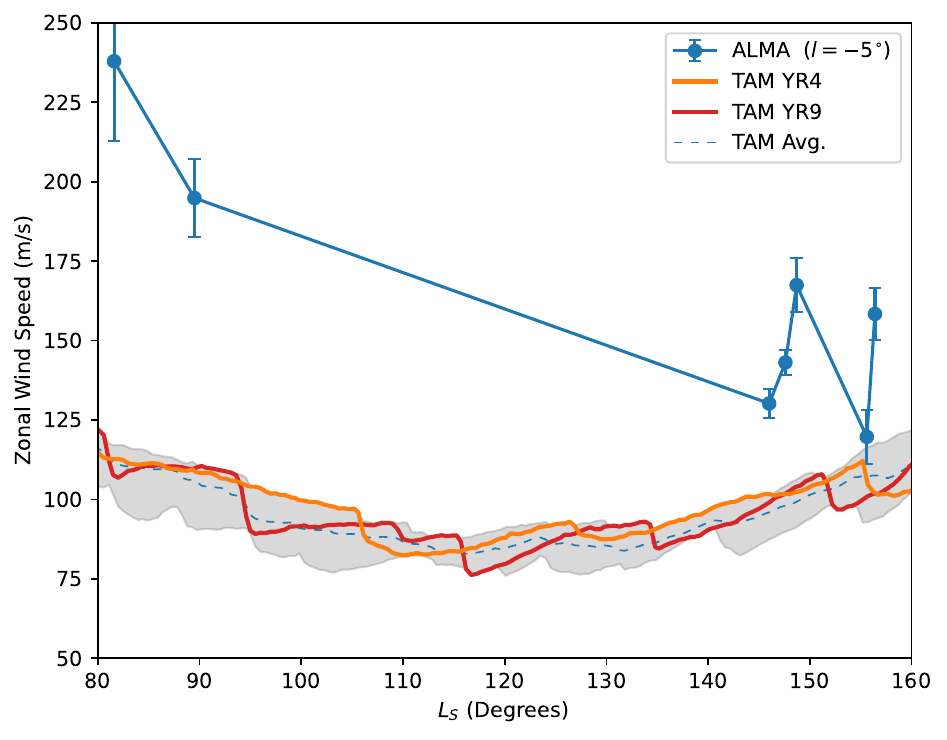}\\
\caption{Observed and theoretical zonal wind velocity profiles as a function of time (expressed in terms of $L_s$), for the period 2022-06-10 to 2023-05-09. Top panel: winds extracted at latitudes $l=-5^{\circ}$ and $l=-45^{\circ}$, corresponding to the equatorial and high-latitude winds, respectively, compared with TitanWRF and TAM model data (averaged over 10 Saturnian years, and vertically weighted using the mean CH$_3$CN contribution function shown in Figure \ref{fig:contrib}). Middle: observed zonal winds at $l=-5^{\circ}$ compared with vertically-averaged TitanWRF output, where the grey envelope shows the range of model wind speeds across 10 model years, and the solid lines are for particular (arbitrarily chosen) model years. Bottom: same as Middle, but using TAM GCM output. Wind speeds predicted by the GCMs are typically lower and less variable than those observed with ALMA, particularly in the vicinity of the equator.\label{fig:windtime}}
\end{figure}

Between 2016-08-19 to 2022-06-10 ($L_s=81.6^{\circ}$--$146.0^{\circ}$), the speed of the low-latitude winds decreased monotonically by a total of $46\pm5$\% (from $240\pm26$~m\,s$^{-1}$ to $130\pm5$~m\,s$^{-1}$), then fluctuated between $\sim130$~m\,s$^{-1}$ and $\sim170$~m\,s$^{-1}$ between 2022-06-10 to 2023-04-12 ($L_s=146.0^{\circ}$--$155.6^{\circ}$),  before undergoing a final, substantial increase from $120\pm8$~m\,s$^{-1}$ to $159\pm8$~m\,s$^{-1}$ ($33\pm10$\%) during the short period of only 27 Earth days (1.7 Titan days) between 2023-04-12 and 2023-05-09 ($L_s=155.6^{\circ}$--$156.4^{\circ}$). Conversely, the speed of the high-latitude winds remained constant (within errors) during the same (latter) 27 day period. Before that, the speed of the high-latitude winds was clearly also variable, but at a level more difficult to quantify due to latitudinal blending with the stronger lobe of the low-latitude (equatorial) winds. The observed, abrupt changes in equatorial (and high-latitude) wind speeds are highlighted in Figure \ref{fig:windtime}. 

\section{Comparison of Observed Zonal Winds with GCM Predictions}

Titan GCMs aim to simulate, and reveal the physics behind, the atmospheric climate and circulation, including zonal winds. For the present study, we compare our results with the vertically-averaged zonal winds as a function of latitude from the TitanWRF \citep{new11,lia25} and TAM \citep{lor15,lom23,lom23b} models. To ensure a meaningful comparison with observations, the TitanWRF and TAM models were run to the point of statistical convergence \citep{lom23b}, although the chaotic/episodic nature of the model instabilities as a function of time means that significant inter-annual variability of the model wind speeds is still apparent. We therefore produced mean wind fields from the GCM output averaged over 10 model years, for comparison with the ALMA observations.

\begin{figure}[h!]
\centering
\includegraphics[width=\columnwidth]{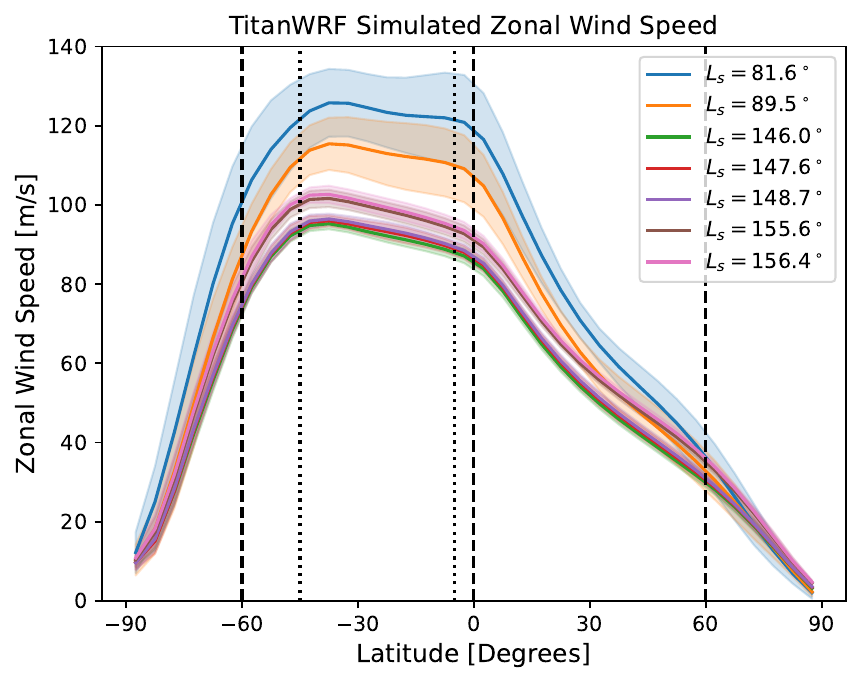}
\includegraphics[width=\columnwidth]{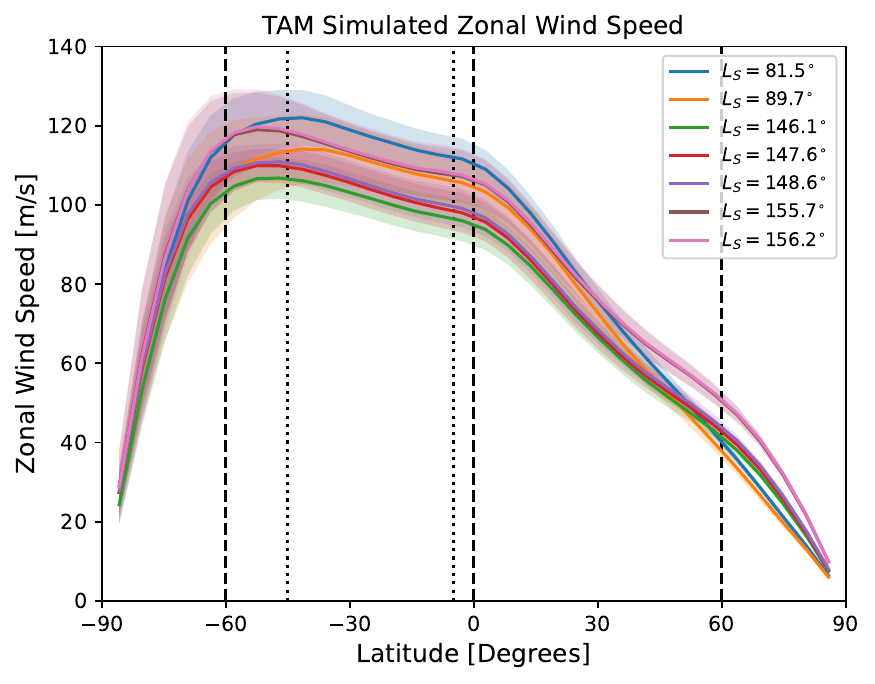}
\caption{Model zonal wind velocity profiles ($v(l)$) for the approximate dates ($L_s$) observed by ALMA, extracted from (left) the TitanWRF GCM \citep{lia25}, and (right) the TAM GCM \citep{lom23b}. Wind speeds have been averaged over 10 model years and vertically weighted using the average CH$_3$CN contribution function (Figure \ref{fig:contrib}). Vertical dashed lines indicate $l=-60^{\circ}$, $0^{\circ}$, and $60^{\circ}$; dotted lines indicate $l=-45^{\circ}$ and $-5^{\circ}$.  $1\sigma$ uncertainty envelopes indicate the standard deviation of the wind speeds over the (multi-year) modeled periods. \label{fig:wrf}}
\end{figure}

The average wind speeds from two representative latitudes from the models (and observations) are shown in Figure \ref{fig:windtime}, with detailed latitudinal profiles (and uncertainty envelopes) as a function of time in Figure \ref{fig:wrf}. These were obtained by averaging the 10-year model wind field as a function of longitude, then applying the average altitudinal weighting function applicable to the ALMA CH$_3$CN wind observations (shown in Appendix B, Figure \ref{fig:contrib}), to match the altitudinal sensitivity. The resulting GCM $v(l)$ profiles show general similarities with those observed using ALMA --- in particular, the presence of strong winds in the mid-to-high-latitude winter hemisphere (around $-30^{\circ}$ to $-50^{\circ}$ south), the relative strengths of which vary with respect to the speed at the equator. The general trend for decreasing peak wind speeds between $L_s=81.6^{\circ}$ to $146^{\circ}$, followed by a subsequent increase towards $L_s=148.7^{\circ}$ is also reproduced by the models.  It should be noted, however, that the quantitative details differ. The modeled wind speeds are faster at high winter latitudes (around $-40^{\circ}$ south) than at the equator, and although the high-latitude model wind speeds are similar to ($\lesssim20$\% lower than) the observations, the simulated equatorial wind speeds are as much as a factor of two slower than the observations.  The sharp ($\sim40$~m\,s$^{-1}$; $\sim27$\%) increase in observed peak (equatorial) wind speeds from $130\pm5$~m\,s$^{-1}$ to $167\pm8$~m\,s$^{-1}$ (over the short time period between $L_s=146.0^\circ$ to $L_s=147.8^\circ$), followed immediately by a decrease of similar magnitude (between $L_s=148.7^\circ$ to $L_s=155.6^\circ$), is also absent from the average model behavior, which shows a more stable (slowly increasing) trend over the same time period, as the zonal winds strengthen towards equinox \citep{lom23b}.

Significant differences between TitanWRF and TAM include the different dynamical cores used to solve the fluid dynamics, and different detailed treatments of heating due to radiative transfer. TAM also includes methane moist convection, condensation, and precipitation \citep{lor15}. Differences between the zonal wind fields produced by TAM and TitanWRF are therefore expected. Despite the model differences and complexities, it is noteworthy that both GCMs produce similar latitudinal zonal wind profiles and temporal trends throughout the observed time period. As shown in Figure \ref{fig:windtime}, for the period between $L_s=146^{\circ}$--$156^{\circ}$, the average TAM wind speeds provide a slightly better match to the overall amplitude of the observations (particularly at high latitudes), although neither model fully reproduces the abruptness of the temporal variations. 

Looking at some of the individual years of the simulations (instead of the 10-year averages), both models show short-timescale wind variability during this period (on a timescale of $\sim1$ Earth year). The amplitude of the variability in some of the individual TAM simulation years is substantial, and more similar to that seen in the later ALMA observations than TitanWRF, which shows less variation over this timeframe, manifested by a relatively narrow envelope of simulated wind values for $L_s>120^{\circ}$ in Figure \ref{fig:windtime} (middle panel). On the other hand, between $L_s=80^{\circ}$--$100^{\circ}$, the $\sim30$~m\,s$^{-1}$ drop in the TitanWRF equatorial zonal wind speeds (for model year 8) is more similar to the observed decrease around that time, compared to that simulated using TAM.

The observed, short-timescale fluctuations between $L_s=146^{\circ}$--$156^{\circ}$ may be related to atmospheric instabilities, which can transport angular momentum and therefore accelerate or decelerate the zonal winds. For instance, \citet{lew23} and \citet{lia25} both found that Rossby-Kelvin instabilities are principally responsible for the equatorial acceleration of superrotation and the mid-to-high latitude deceleration in the stratosphere. While neither TAM nor TitanWRF captures rapid increases in the zonal wind on a time scale of $\Delta{L_s}\sim 5^{\circ}$ near $-5^\circ$,  they both show evidence for fast drops in the equatorial wind speed near $L_s\sim 150^\circ$, albeit only for particular simulation years (as shown in Figure \ref{fig:windtime}), highlighting the variable nature of the process. We suggest that the abrupt nature of this drop could be due to temporary weakening of Rossby-Kelvin instabilities, leading to equatorial deceleration. To understand the mismatch between the observations and the models (particularly the rapid acceleration observed by ALMA) and the different behavior between TAM and TitanWRF from year to year (as well as the overall wind acceleration/deceleration trend in both models between $L_s=80^\circ$ to $L_s=160^\circ$),  a detailed analysis of wave-mean flow interactions during this time period is needed for both models, which will be deferred to future work.

\begin{figure*}[t!]
\centering
\includegraphics[width=0.6\textwidth]{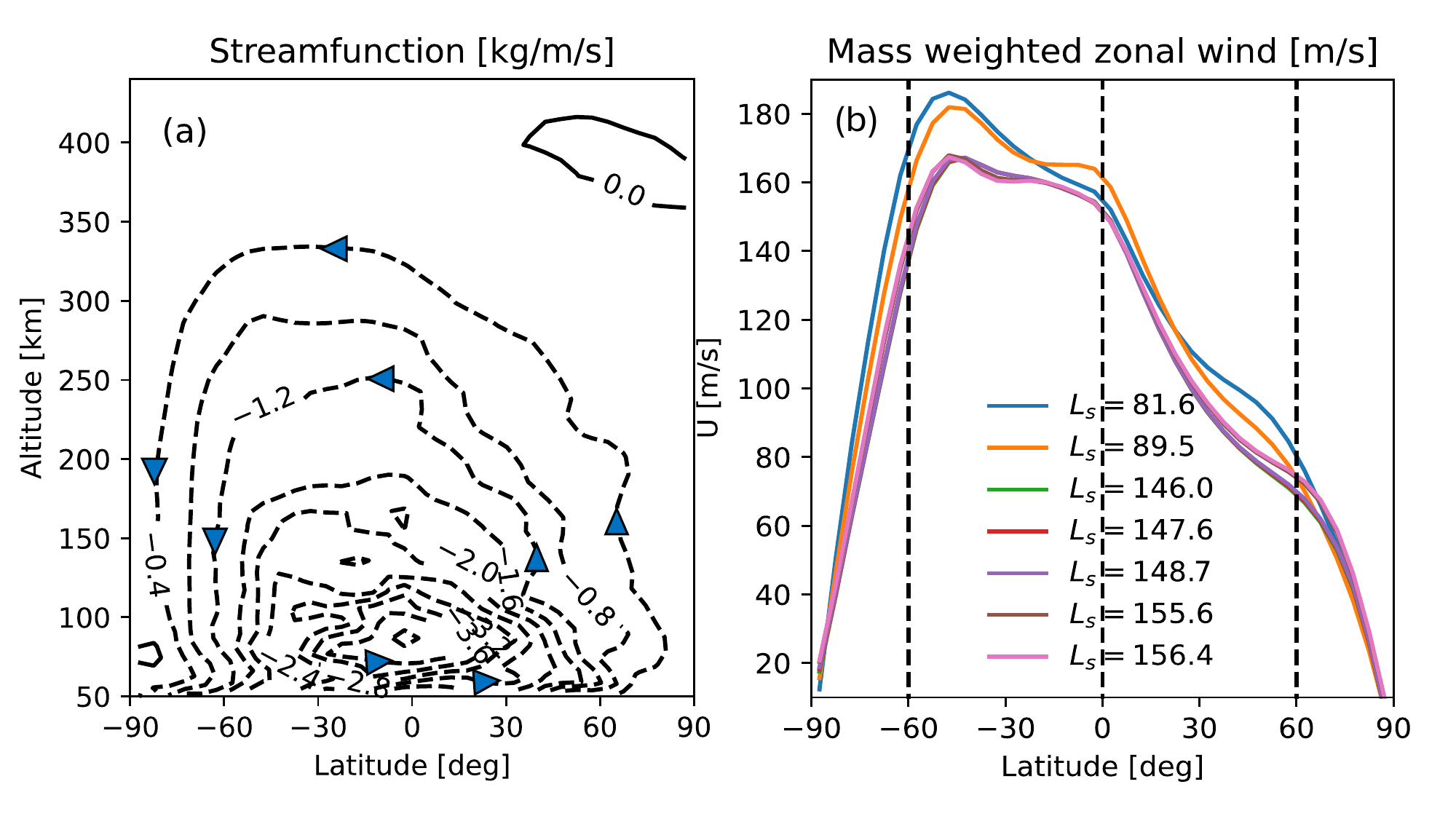}
\caption{TitanWRF zonal mean streamfunction (a) and mass-weighted zonal winds (b). The streamfunction is time-averaged between $L_s=81.6^{\circ}$ and $L_s=156.4^{\circ}$, with the solid lines (non-negative values) for clockwise rotation and the dashed lines (negative values) for counter-clockwise rotation (as depicted by the arrows). The mean zonal wind is mass-weighted between 150 km ($\sim2$ mbar) and 450km ($\sim 5$~nbar) at the selected $L_s$ corresponding to the timing of the ALMA observations. Note that the wind profiles from $L_s=146^{\circ}$ to $L_s=156.4^{\circ}$ are mostly overlapped.
\label{fig:wrf_psi_u}}
\end{figure*}

Interpretation of the time evolution of the zonal wind field is facilitated by considering the difference {in the vertical profile shapes of the CH$_3$CN contribution function and Titan's air density function. The former provides insight into the specific (range of) altitudes to which our ALMA observations are most sensitive, whereas} the latter is more appropriate for diagnosing the bulk wind evolution in terms of conservation of angular momentum. For example, both the TitanWRF model results and the ALMA observations (weighted by the CH$_3$CN vertical contribution function) show a noticeable decrease of zonal wind between $\pm 50^\circ$ from $L_s=81.6^\circ$ to $L_s=89.5^\circ$. On the other hand, the mass-weighted wind $U_m = (\int_{z_b}^{z_t} \rho u \,dz)/(\int_{z_b}^{z_t} \rho \,dz)$ in the same altitude range ($z_{b}=150$~km and $z_t=450$~km) shows an increase of wind speed at the equator and decreases of wind speed at mid-latitudes in both hemispheres (Figure \ref{fig:wrf_psi_u}b). Here $\rho$ is the air density, $u$ is the zonal wind speed and $dz$ is the layer thickness. {This discrepancy arises because the mass-weighted average is heavily biased by the denser, lower atmosphere ($z \sim 150$~km), whereas the CH$_3$CN contribution function is most sensitive to higher altitudes (peaking around $z \sim 250$~km). Because the higher-altitude (CF-weighted) equatorial winds are seen to decelerate while the lower-altitude (mass-weighted) winds simultaneously accelerate}, we can infer a net downward transport of equatorial angular momentum from higher to lower altitudes during the period around northern summer solstice. This is also consistent with the behavior of the main zonal jet system seen in TAM, which descends from its altitudinal peak in the mesosphere around equinox to the stratosphere around solstice, accompanied by a slowing of the higher-latitude zonal winds \citep{lom23b}.

The zonal wind in Titan's middle atmosphere is primarily driven by the advection of absolute vorticity, the vertical advection of momentum, and the horizontal/vertical transport of momentum by eddies. Their relative importance depends on the locations of the atmosphere and seasons \citep{lom23b}. The mass-weighted zonal wind shows noticeable temporal variations during the first three recorded solar longitudes $L_s=81.6^\circ$, $L_s=89.5^\circ$ and  $L_s=146.0^\circ$, while it remains nearly unchanged at other solar longitudes in the altitude range between $z_{b}=150$ km and $z_t=450$ km.  From  $L_s=81.6^\circ$ to $L_s=89.5^\circ$, the zonal wind decreases in the northern and southern mid-to-high latitudes over time, but it is slightly accelerated at the low latitudes between $-20^\circ$ and $20^\circ$. The deceleration at the northern latitudes poleward of $20^\circ$ is mainly a result of vorticity advection by the meridional flow (Figure \ref{fig:wrf_psi_u}a), while the deceleration at the southern latitudes (poleward of $-20^\circ$) is a result of vertical momentum advection and transport of momentum by horizontal eddies, both of which counteract the acceleration caused by the poleward meridional flow. The acceleration of the equatorial wind between $-20^\circ$ and $20^\circ$ results from the deposition of momentum there by the horizontal eddies \citep{lew23,lia25}. 

Although the detailed wind speeds differ, the TitanWRF and TAM simulations show comparable qualitative similarities and differences with respect to the winds retrieved from observations. Considering the dynamical complexity of Titan's atmosphere, the qualitative agreement between the observed and modeled zonal wind fields as a function of time demonstrates that GCMs can simulate reasonably well the atmospheric processes related to Titan's superrotation and its seasonal variation. The major similarities between the TitanWRF and TAM zonal wind fields show that Titan's winds are not strongly affected by moist processes at the altitudes probed by ALMA CH$_3$CN observations. However, the inability to adequately reproduce the full amplitude of the equatorial wind speed (which increases toward higher altitudes; \citealt{lel19,cor20}) hints at a possible avenue towards improving the ability of GCMs to more fully reproduce submillimeter interferometric observations. Due to the close dynamical coupling between atmospheric layers (from tropopause to thermosphere), extending the GCM model top from its present upper limit (around $z\sim500$~km) to thermospheric altitudes ($z\sim1000$~km) would allow the dynamical origin, properties, and ultimate impact of the thermospheric equatorial jet \citep{lel19} to be investigated. We hypothesize that vertical transport of angular momentum between the stratospheric and thermospheric jets could play a role in accelerating the strong equatorial winds observed by ALMA at lower (stratospheric-mesospheric) altitudes. Improved vertical grid resolution may also allow GCMs to more fully capture the horizontal and vertical angular momentum transport mechanisms associated with equatorial waves \citep{lew23}.

\section{Conclusion}

To improve our understanding of Titan's high-altitude atmospheric circulation, we have performed ALMA Doppler mapping of the zonal wind field using new CH$_3$CN observations from the period 2022-2023, combined with archival data from 2016-2017 to cover the period between southern winter and spring. The resulting time-series of latitudinal wind profiles is primarily sensitive to altitudes in the upper stratosphere (in the range $z\approx200$-490~km), with improved equatorial coverage compared with previous thermal wind speed determinations based on Cassini CIRS data.

Our ALMA observations reveal a trend for consistently time-variable zonal winds over a broad range of latitudes. Particular features of our analysis include the presence of a strongly-variable equatorial jet, combined with more weakly-variable winds at mid-to-high southern (winter) latitudes. For the first time, our ALMA data reveal strong ($30$--40~m\,s$^{-1}$), ongoing temporal variability in Titan's wind speeds at equatorial (and higher) latitudes, over timescales as short as $\approx1$ Earth month ($0.9^{\circ}$ in $L_s$) or less. 

We compare the observed temporal variability of the zonal winds with simulations using the TitanWRF and TAM GCMs. Our results show that the evolution of Titan's equatorial and high-latitude middle atmospheric zonal winds are adequately reproduced by the models at seasonal timescales, indicating that the physical processes captured in these simulations are robust. Variations in the upper-stratospheric zonal wind speeds during the observed period can occur as a result of horizontal (meridional) and vertical (altitudinal) transport of angular momentum, mediated by meridional advection and Rossby-Kelvin instabilities. The models provide a reasonable qualitative explanation for the broader (seasonal) trends in the observed wind speed behavior between $L_s=81.6^{\circ}$--$156.4^{\circ}$. However, the detailed latitudinal wind speed profiles, their precise temporal variability, and the absolute magnitudes of the peak wind speeds are not well reproduced, so further theoretical studies are needed.

Compared to previous zonal wind measurements based on analysis of Cassini infrared temperature data, our ALMA observations are uniquely sensitive to wind speeds in the vicinity of Titan's equator, leading to a more complete picture of the global atmospheric dynamics. Within the limitations of our latitudinal resolution ($\sim30^{\circ}$), we find that the fastest wind speeds consistently occur within $-10^{\circ}$ of the equator. The TitanWRF or TAM GCMs, on the other hand, exhibit fastest winds at higher (winter) latitudes (around $-40^{\circ}$ to $-50^{\circ}$ south). Improvements to the models to help better simulate upper-level equatorial winds could include raising the model top and adopting a finer vertical grid spacing to better capture high-altitude equatorial phenomena and wave-propagation effects. Combined with detailed analysis of the energy and momentum transport processes in the models, this will facilitate a more complete interpretation of the observations, and shed further light on the unexpectedly rapid variations in Titan's zonal jets. Continued temporal monitoring with ALMA will also be needed (including at higher spatial and temporal resolution, over a range of altitudes), in order to more fully characterize Titan's global wind field across seasons, thus providing further benchmarks for future GCM efforts, toward a refined understanding of Titan's atmospheric dynamics.

\begin{acknowledgments}
This work was supported by the NASA ROSES Solar System Observations program, the National Science Foundation under grant No. AST-2407709, and NASA’s Planetary Science Division Internal Scientist Funding Program through the Fundamental Laboratory Research work package (FLaRe). NAT and LW are funded by UK Science and Technology Facilities Council grant ST/Y000676/1.
This work makes use of ALMA data sets  ADS/JAO.ALMA\#2015.1.01023.S, \#2016.A.00014.S, \#2021.A.00034.S and \#2021.1.00736.S. ALMA is a partnership of ESO, NSF, NINS, NRC, NSC and ASIAA, in cooperation with the Republic of Chile. The Joint ALMA Observatory is operated by ESO, AUI/NRAO and NAOJ.
\end{acknowledgments}

\clearpage

\bibliography{TitanRefs}
\bibliographystyle{aasjournal}

\software{CASA (v6.5; \citealt{casa22}), AtmWindMapGaussFit \citep{cor20}, LMFIT \citep{new16}.}

\appendix

\section{Doppler Wind Error Maps}

\begin{figure}
\begin{center}
 \includegraphics[width=0.4\columnwidth]{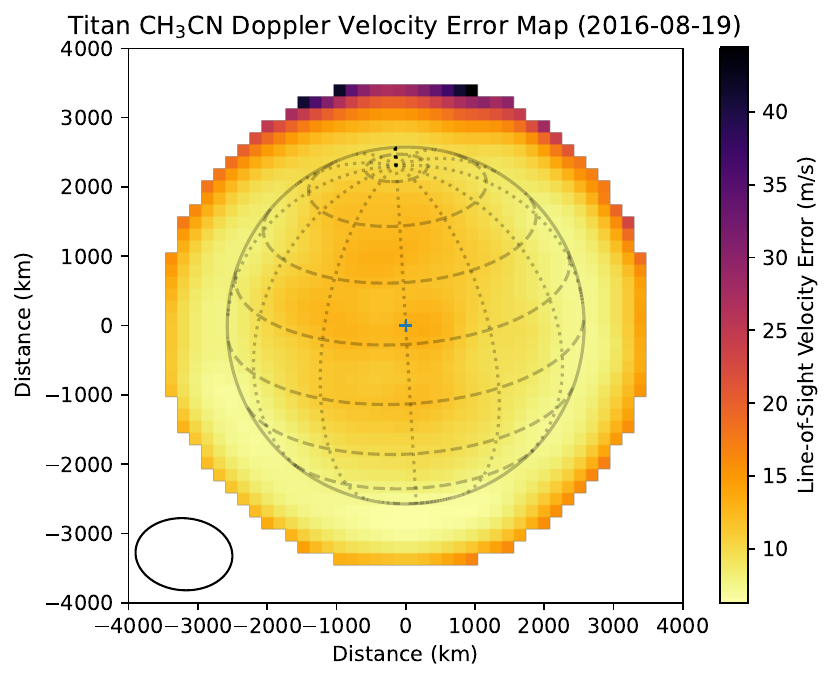}\\
 \includegraphics[width=0.4\columnwidth]{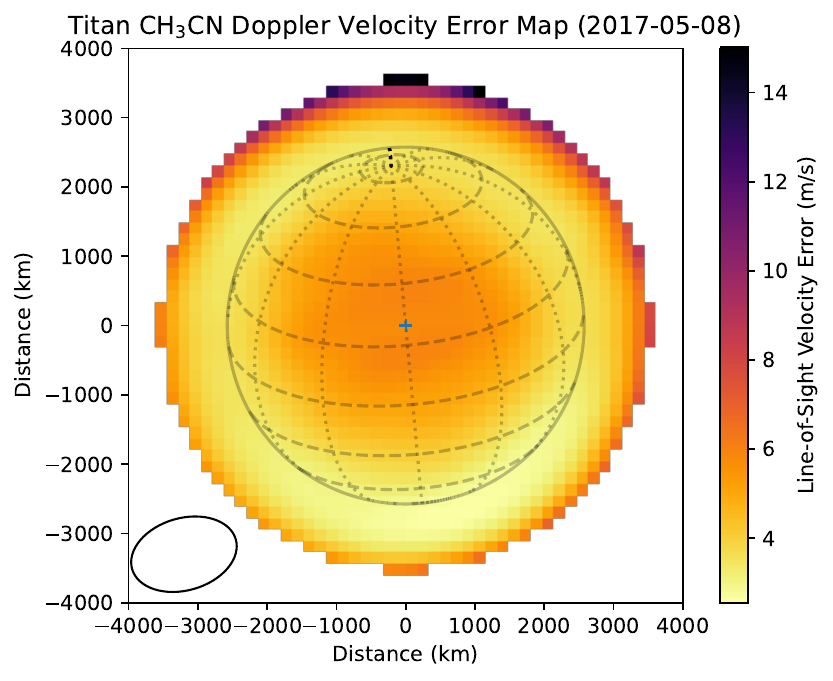}\\
 \includegraphics[width=0.4\columnwidth]{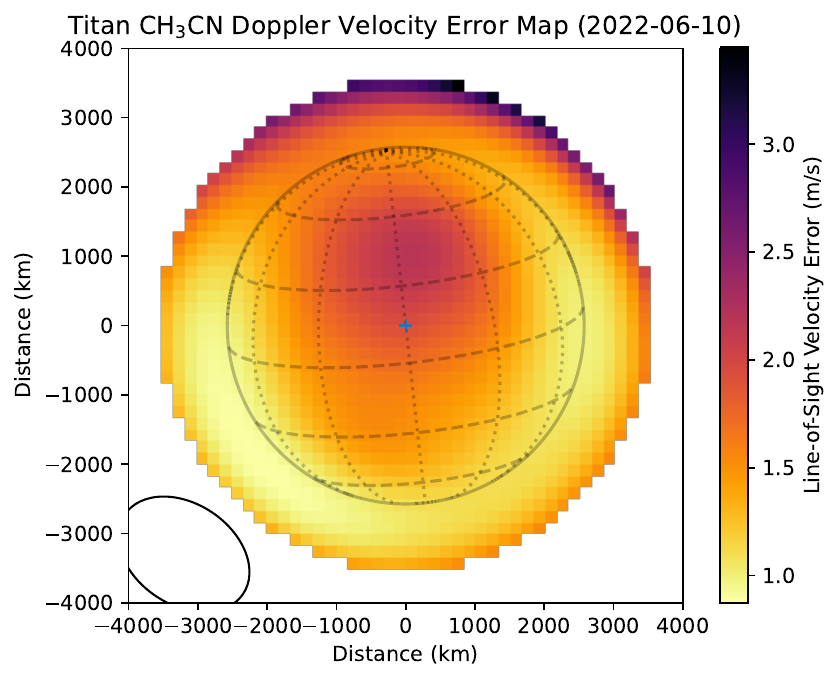}\\
 \includegraphics[width=0.4\columnwidth]{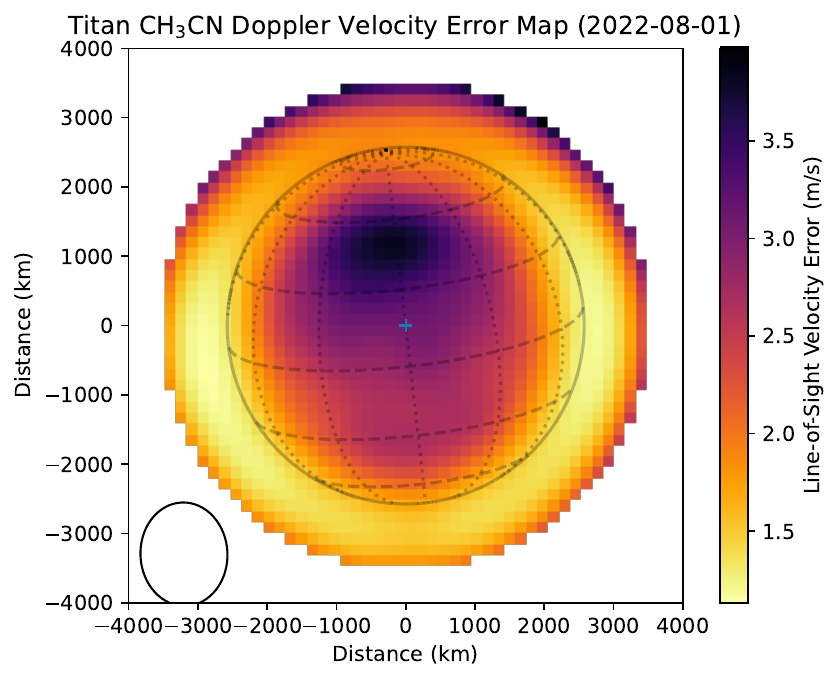}\\
 \caption{Color maps showing statistical uncertainties on the line-of-sight CH$_3$CN Doppler shifts, for four epochs (2016-08-19 to 2022-08-01). Titan's wireframe grid is shown for each epoch, with the spatial resolution (elliptical beam FWHM) indicated lower left.}
   \label{fig:errmaps1}
\end{center}
\end{figure}

\begin{figure}
\begin{center}
 \includegraphics[width=0.4\columnwidth]{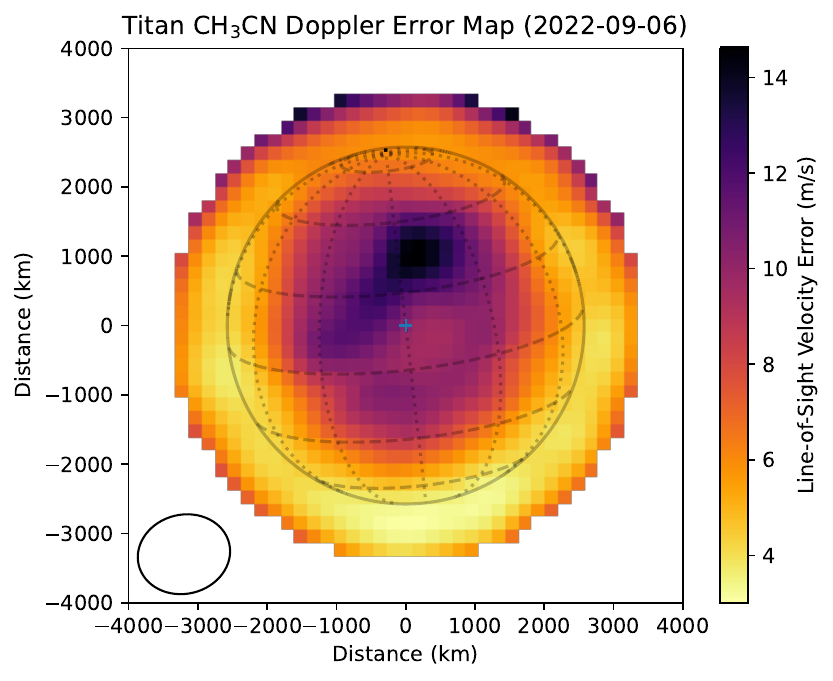}\\
 \includegraphics[width=0.4\columnwidth]{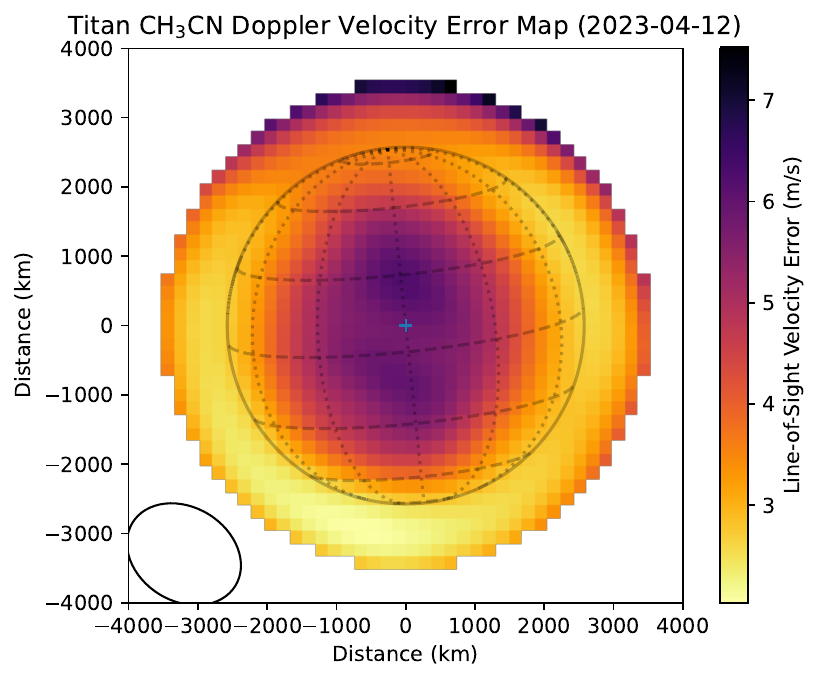}\\
 \includegraphics[width=0.4\columnwidth]{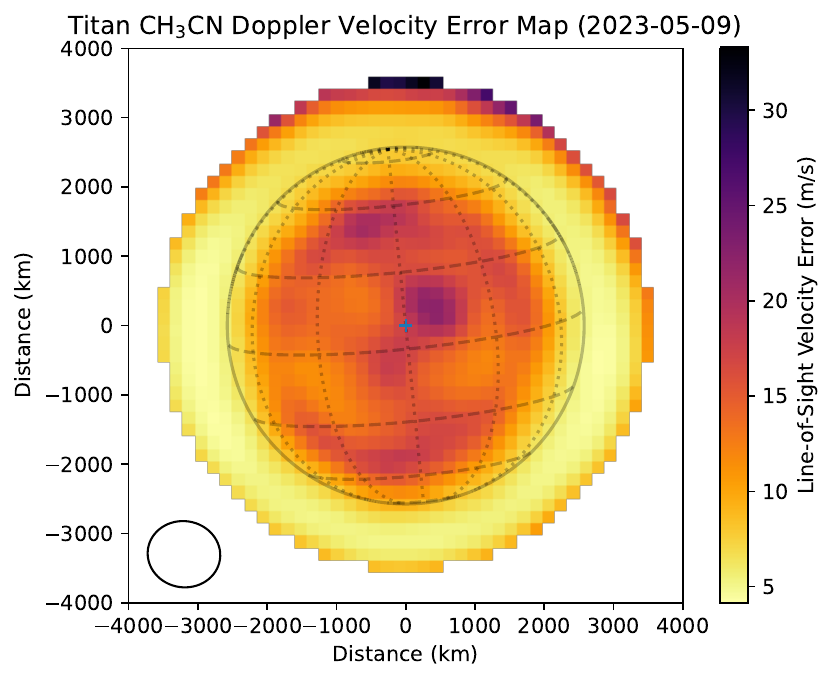}\\
 \caption{Same as Figure \ref{fig:errmaps1}, for epochs 2022-09-06 to 2023-05-09.}
   \label{fig:errmaps2}
\end{center}
\end{figure}

For each ALMA epoch, we performed least-squares fits to the observed CH$_3$CN spectral lines using a combination of Moffat functions, to derive the gas Doppler shift (along the line of sight) in each image pixel (see Section \ref{sec:fits}). Uncertainties on the Doppler shifts were determined from the diagonal elements of the covariance matrix of fit parameters, and are plotted as error maps in Figures \ref{fig:errmaps1} and \ref{fig:errmaps2}.

\section{CH$_3$CN Vertical Abundance Profiles and Contribution Functions}

\begin{figure}[h!]
\centering
\includegraphics[width=0.6\columnwidth]{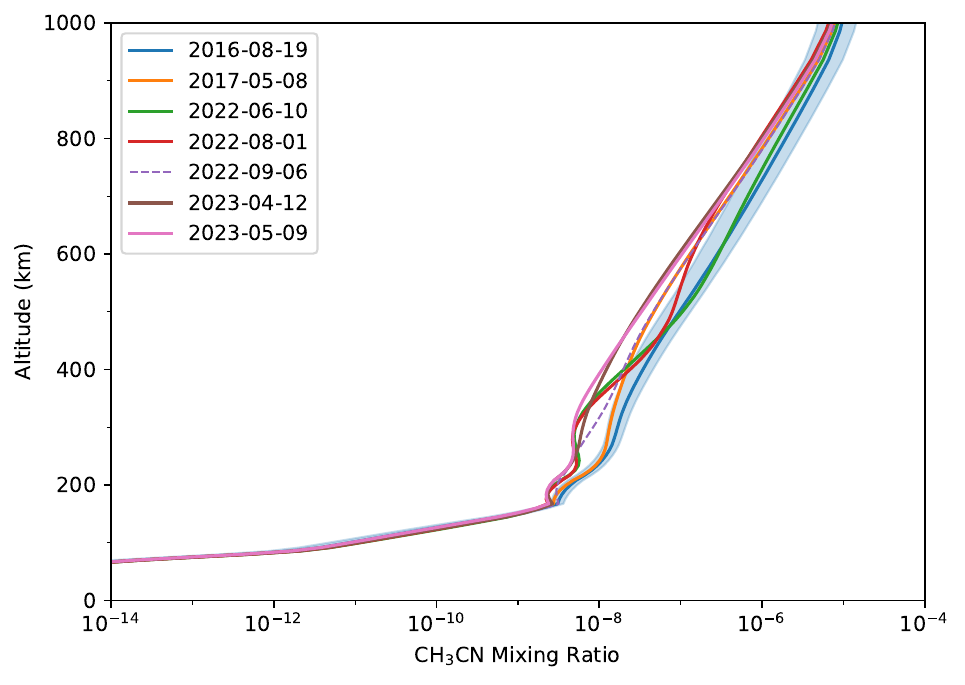}
\caption{Retrieved CH$_3$CN vertical abundance profiles for each epoch. To illustrate the approximate size of the uncertainties, the NEMESIS retrieval error envelope is shown in blue, for the 2016-08-19 epoch. The 2022-09-06 retrieval is shown with a dashed linestyle to indicate additional (unquantified) uncertainty as a result of extra line broadening introduced by the rubidium maser problem identified in \ref{sec:fits}.   \label{fig:vmr}}
\end{figure}

\begin{figure}[h!]
\centering
\includegraphics[width=0.5\columnwidth]{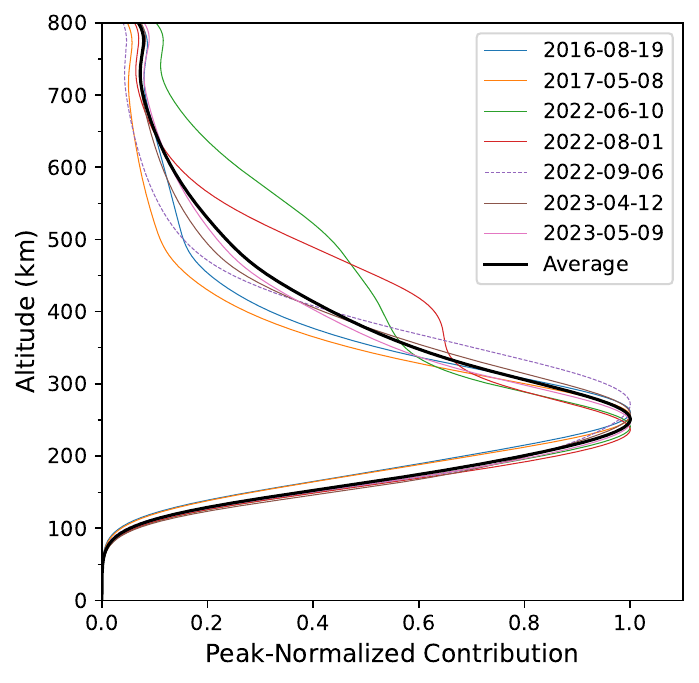}
\caption{Peak-normalized wind contribution functions for each epoch, derived from the NEMESIS models in Figure \ref{fig:vmr}. The mean contribution function profile is shown in black. \label{fig:contrib}}
\end{figure}

CH$_3$CN mixing ratios as a function of altitude were retrieved using NEMESIS based on radiative transfer model fits to the observed spectral line profiles, and are shown for each epoch in Figure \ref{fig:vmr}. The associated wind contribution functions are shown in Figure \ref{fig:contrib}, calculated by multiplying the functional derivatives ($dQ/d\log(x)$)$_{\nu}$ of the limb emission by the gradient of the spectral line model $dQ/d{\nu}$, then averaging over all model frequencies $\nu$  (see Section \ref{sec:fits}).

\end{document}